\documentclass[sigconf]{acmart}
\usepackage{xcolor}
\newcommand{\bw}[1]{\textcolor{black}{#1}}
\AtBeginDocument{%
  }
\usepackage{listings}

\copyrightyear{2023}
\acmYear{2023}
\setcopyright{rightsretained}
\acmConference[CHI '23]{Proceedings of the 2023 CHI Conference on Human Factors in Computing Systems}{April 23--28, 2023}{Hamburg, Germany}
\acmBooktitle{Proceedings of the 2023 CHI Conference on Human Factors in Computing Systems (CHI '23), April 23--28, 2023, Hamburg, Germany}
\acmDOI{10.1145/3544548.3580895}
\acmISBN{978-1-4503-9421-5/23/04}

\usepackage{listings}
\usepackage{xcolor}

\definecolor{codegreen}{rgb}{0,0.6,0}
\definecolor{codegray}{rgb}{0.5,0.5,0.5}
\definecolor{codepurple}{rgb}{0.58,0,0.82}
\definecolor{backcolour}{rgb}{0.95,0.95,0.92}

\lstdefinestyle{mystyle}{
    backgroundcolor=\color{backcolour},   
    commentstyle=\color{codegreen},
    keywordstyle=\color{magenta},
    numberstyle=\tiny\color{codegray},
    stringstyle=\color{codepurple},
    basicstyle=\ttfamily\footnotesize,
    breakatwhitespace=false,         
    breaklines=true,                 
    captionpos=b,                    
    keepspaces=true,                 
    numbers=left,                    
    numbersep=5pt,                  
    showspaces=false,                
    showstringspaces=false,
    showtabs=false,                  
    tabsize=2,
}

\begin{document}

\title[Enabling Conversational Interaction with Mobile UI using Large Language Models]{Enabling Conversational Interaction with Mobile UI using \\ Large Language Models}




\author{Bryan Wang}
\authornote{Work done during an internship at Google Research.}
\affiliation{%
  \institution{University of Toronto}
  \city{Toronto}
  \state{ON}
  \country{Canada}}
\email{bryanw@dgp.toronto.edu} 

\author{Gang Li}
\affiliation{%
  \institution{Google Research}
  \city{Mountain View}
  \state{CA}
  \country{USA}}
\email{leebird@google.com} 

\author{Yang Li}
\affiliation{%
  \institution{Google Research}
  \city{Mountain View}
  \state{CA}
  \country{USA}}
\email{liyang@google.com} 
\renewcommand{\shortauthors}{Wang et al.}

\begin{abstract}
Conversational agents show the promise to allow users to interact with mobile devices using language. However, to perform diverse UI tasks with natural language, developers typically need to create separate datasets and models for each specific task, which is expensive and effort-consuming. Recently, pre-trained large language models (LLMs) have been shown capable of generalizing to various downstream tasks when prompted with a handful of examples from the target task. This paper investigates the feasibility of enabling versatile conversational interactions with mobile UIs using a single LLM. We designed prompting techniques to adapt an LLM to mobile UIs. We experimented with four important modeling tasks that address various scenarios in conversational interaction. Our method achieved competitive performance on these challenging tasks without requiring dedicated datasets and training, offering a lightweight and generalizable approach to enable language-based mobile interaction.


\end{abstract}



\maketitle

\section{Introduction}

Interacting with computing devices using natural language is a long-standing pursuit in human-computer interaction \cite{bolt1980put, karat2002conversational, folstad2017chatbots}. Language, as both the input and the output, allows users to efficiently communicate with a computing system and access its functionalities when other I/O modalities are unavailable or cumbersome. The interaction paradigm is particularly useful for users with motor or visual impairments or situationally impaired when occupied by real-world tasks \cite{wobbrock2019situationally, sarsenbayeva2017challenges}. 
Intelligent assistants, e.g., Google Assistants and Siri, have significantly advanced language-based interaction for performing simple daily tasks such as setting a timer. 
Despite the progress, these assistants still face limitations in supporting conversational interaction in mobile UIs, where many user tasks are performed \cite{li2017sugilite}. For example, they cannot answer a user's question about specific information displayed on the screen \cite{screenqa}. Achieving such capabilities requires an agent to have a computational understanding of graphical user interfaces (GUIs), which is absent in existing assistants. 

Prior research has investigated several important technical building blocks to enable conversational interaction with mobile UIs, including summarizing a mobile screen for users to quickly understand its purpose \cite{10.1145/3472749.3474765}, mapping language instructions to UI actions \cite{li-etal-2020-mapping, pasupat-etal-2018-mapping, liu2018reinforcement} and modeling GUIs so that they are more amenable for language-based interaction \cite{wu2021screen, li2021screen2vec, zhang2021screen, li2021vut, 10.1145/3472749.3474765}. However, each of them only addresses a limited aspect of conversational interaction and requires considerable effort in curating large-scale datasets and training dedicated models. \cite{li-etal-2020-mapping, 10.1145/3472749.3474765, vut2021, li2020widget}. Furthermore, there is a broad spectrum of conversational interactions that can occur on mobile UIs, as Todi et al. revealed~\cite{todi21conversations}. Therefore, it is imperative to develop a lightweight and generalizable approach to realize conversational interaction.

Recently, pre-trained large language models (LLMs) such as GPT-3 \cite{brown2020language} and PaLM \cite{chowdhery2022palm} have demonstrated abilities to adapt themselves to various downstream tasks when being \textit{prompted} with a handful of examples of the target task. Such generalizability is promising to support diverse conversational interactions without requiring task-specific models and datasets. However, the feasibility of doing so is unclear. Little work has been conducted to understand how LLMs, trained with natural languages, can be adapted to GUIs for interaction tasks. Therefore, we investigate in this paper the viability and the how-to of utilizing LLMs to enable diverse language-based interactions with mobile UIs. 

 We categorized four mobile UI conversational interaction scenarios, which guided our experimental task selection. We developed a set of prompting techniques to prompt LLMs with mobile UIs. Since LLMs only take text tokens as input, we contribute an algorithm to generate the text representation of mobile UIs. Our algorithm uses depth-first search traversal to convert the Android UI's view hierarchy, i.e., the structural data containing detailed properties of UI elements, into HTML syntax. We conducted comprehensive experiments with four pivotal modeling tasks, including \textit{Screen Question-Generation}, \textit{Screen Summarization}, \textit{Screen Question-Answering}, and \textit{Mapping Instruction to UI Action}. The experimental results show that our approach achieves competitive performance using only \textit{two data examples} per task. Notably, our work is the first to investigate methods to enable the \textit{Screen Question-Generation} and \textit{Screen Question-Answering} tasks in literature, setting benchmark performances. Furthermore, the human evaluation of the \textit{Screen Summarization} task demonstrates that our method generates more accurate screen summaries compared to Screen2Words \cite{10.1145/3472749.3474765}, the benchmark model trained with tens of thousands of examples. In summary, the evaluation results from a suite of modeling tasks validate the feasibility and effectiveness of our approach in enabling conversational interaction with mobile UIs.

More broadly, our study demonstrates LLMs' potential to fundamentally transform the future workflow of conversational interaction design. Using our prompting method, interaction designers and developers can quickly prototype and test novel language interactions with users, saving time and resources before investing in dedicated datasets and models. Our experiments are based on open-source Android UI datasets, including RICO \cite{Deka:2017:Rico}, Screen2Words \cite{10.1145/3472749.3474765}, and PixelHelp \cite{li-etal-2020-mapping}. We also open-source the code \footnote{\url{https://github.com/google-research/google-research/tree/master/llm4mobile}} of our algorithm that converts the Android view hierarchy to HTML syntax to allow future work to replicate and build upon our work. Example prompts created using our algorithm can be found in the appendix. In summary, our paper makes the following contributions:
\begin{itemize}
    \item Our work is the first investigation for using LLMs to enable conversational interaction on mobile UIs, which advances the understanding of using LLMs for interaction tasks.
    \item We designed a novel method for feeding GUIs to LLMs---that is pre-trained for natural language---and a set of techniques to prompt LLMs to perform a range of conversational tasks on mobile UI screens. These techniques produce competitive performance; \bw{we open-source the code} so others can immediately use them in their work.
    \item We experimented with four pivotal modeling tasks, demonstrating the feasibility of our approach in adapting LLMs for conversational GUI interaction and potentially lowering the barriers to developing conversational agents for GUIs.
\end{itemize}

\section{Related Work}
\subsection{Bridging GUIs with Natural Language}
There has been increasing interest in using machine learning to bridge graphical user interfaces and natural language for use cases such as accessibility and multimodal interaction. For example, Widget Captioning \cite{li2020widget} and Screen Recognition  \cite{zhang2021screen} predict semantically meaningful alt-text labels for GUI components. Screen2Words \cite{10.1145/3472749.3474765} took a step further to predict text summaries that concisely describe the entire screen using multimodal learning. Leiva et al. proposed a vision-only approach to generate templated language descriptions of UI screenshots \cite{10.1145/3564702}. Li et al. \cite{li-etal-2020-mapping} uses a transformer-based model to map natural language instructions to mobile UI action sequences. These prior works typically train a model dedicated to the task based on a sizeable dataset collected. In contrast, our work leverages the few-shot learning ability of LLMs to enable language-based UI tasks by providing a small number of examples. To achieve this, we propose a novel method to represent the UI so that an LLM pre-trained for natural language can efficiently process it. Another relevant body of work is to develop a conversational or multimodal agent that can help the user accomplish mobile tasks \cite{li2017sugilite, li2019pumice, li2020multi, 8506506}. For example, SUGILITE \cite{li2017sugilite} enables users to create task automation on smartphones by user demonstration and perform the tasks through a conversational interface. KITE \cite{8506506} helps developers create task-oriented bots templates from existing apps. Our work shows that LLMs can enable versatile language-based interactions when prompted with exemplars for different tasks, lowering the threshold for developing versatile multimodal agents.

\subsection{Prompting Pre-trained Large Language Models}
Finetuning pre-trained task-invariant models such as BERT has been a common practice to adapt large models for specific tasks. However, GPT-3 \cite{brown2020language} introduced a new norm for leveraging pre-trained language models for downstream tasks through in-context few-show learning, i.e., prompting. By prompting a pre-trained model with only a few examples, it can generalize to various tasks without updating the parameters in the underlying model. Recently studies have shown that prompting is one of the emergent abilities that appear only when the model size is large enough \cite{wei2022emergent}. While prompting LLMs may not always outperform benchmark models, it provides a lightweight method to achieve competitive performance on various tasks \cite{brown2020language, chowdhery2022palm}. When the prompt consists of $N$ pairs of input and output exemplars from the target tasks, it is referred to as $N$-shot learning, and providing more shots typically improves performance \cite{brown2020language, chowdhery2022palm}. Additionally, various prompting paradigms have been proposed to elicit logical reasoning from the language model \cite{wei2022chainofthought, zhou2022leasttomost, wang2022rae, kojima2022think}, which is useful for tasks like solving math problems. For example, Chain-of-Thought prompting \cite{wei2022chainofthought} proposes to use the models to generate intermediate results (i.e., a chain of thoughts) before generating the final output. The core idea resembles the divide-and-conquer method in algorithms, which breaks more complicated problems into subproblems that can be solved more easily. Prompting LLMs remains an ongoing research topic in the community. Our work builds upon prior work to contribute a set of prompting techniques designed to adapt LLMs to mobile UIs.

\begin{figure*}[ht]
  \centering
  \includegraphics[width=\linewidth]{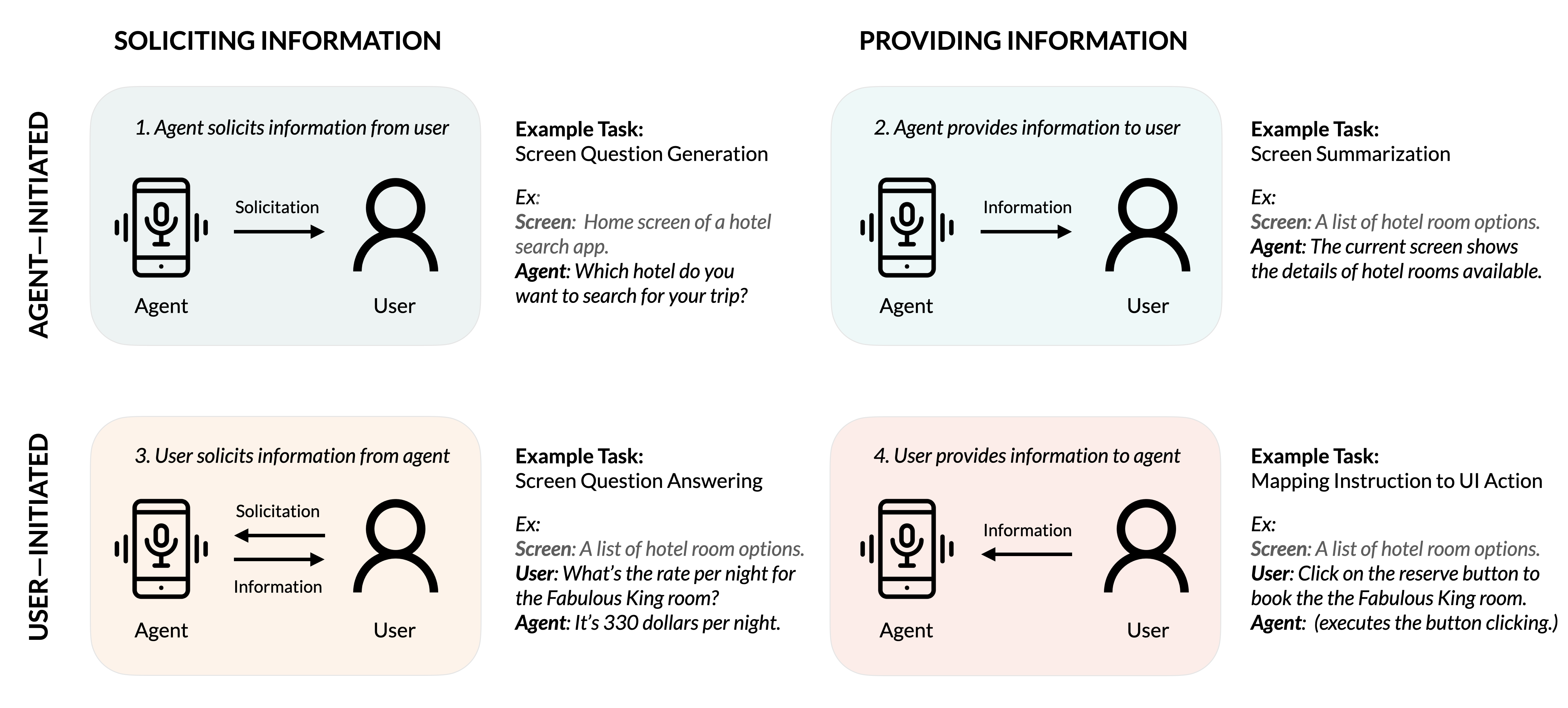}
  \caption{The categorization of different conversation scenarios when user and agent interact to complete tasks on mobile UIs. The categorization has two axes: initiative and purpose. An agent can take the initiative to solicit new information from the user or to provide information to the user, and vice versa. We included an example task for each category where the agent's action represents the target output from LLMs.}   
  \Description{The categorization of different types of unit conversation when the user and the agent work collaboratively toward completing user goals on mobile UIs. The categorization has two axes: initiative and purpose. An agent can take the initiative to solicit new information from the user or to provide information to the user and vice versa. Each category features an example task with the desired agent action highlighted. 1. Agent solicits information from user: Screen Question Generation. 2. Agent provides information to user: Screen Summarization. 3: User solicits information from the agent: Screen Question-Answering. 4. User provides information to the agent: Mapping instruction to UI Action.}
  \label{dspace}
\end{figure*}

\subsection{Interactive Applications of Large Language Models}
LLMs have been applied to enable a broad range of language-related interactive applications in the HCI community \cite{chung2022talebrush, dang2022prompt, lee2022promptiverse, kim2022stylette, 2022saycan, lee2022coauthor, lee2022interactive, wu2022aichain, jian2022case, jiang2022nlp, liu2022will}. For example, Chang et al. \cite{chung2022talebrush} proposed TaleBrush, a generative story ideation tool that uses line sketching interactions with a GPT-based language model for control and sensemaking of a protagonist’s fortune in co-created stories. Stylette \cite{kim2022stylette} allows users to modify web designs with language commands and uses LLMs to infer the corresponding CSS properties. Lee et al. \cite{lee2022coauthor} present CoAuthor, a dataset designed to reveal GPT-3’s capabilities in assisting creative and argumentative writing. Since LLMs can encode a wealth of semantic knowledge, they have also been used to support physical applications. For example, SayCan \cite{2022saycan} extracts and leverages the knowledge priors within LLMs to execute real-world, abstract, long-horizon robot commands. Our work represents the first contribution of applying LLMs to enable conversational interactions on mobile UIs.

\section{Conversation for Mobile UI Tasks} 
\label{conversation_space}
Conversational interaction with mobile devices is typically embodied as human users exchanging information with a conversational agent. We developed a conceptual framework categorizing four conversation scenarios between users and agents when performing mobile tasks. Our categorization has two dimensions: \textit{Initiative} and \textit{Purpose}.  As shown in Figure \ref{dspace}, a conversation can be mixed-initiative \cite{horvitz1999principles}, either initiated by the agent or the user. The purpose of initiated conversation can be either soliciting or providing information. The categorization lays the foundation for determining important modeling tasks to investigate. We focus on studying how LLMs can enable the interaction capabilities of a conversational agent based on a mobile UI, e.g., providing language responses or performing UI actions on behalf of the user. For simplicity, we limit our study to unit conversations which include, at most, a single turn from the user and agent. More complex, multi-turn conversations are beyond the scope of this paper and will be explored in future research, which we discuss in section \ref{multiturn}. We now introduce each conversation category and the associated example modeling tasks, including 1) \textit{Screen Question-Generation}, 2) \textit{Screen Summarization}, 3) \textit{Screen Question-Answering}, and 4) \textit{Mapping Instruction to UI Action}. 

\subsection{Agent-initiated Conversation}
When an agent initiates a conversation, it can be either soliciting or providing information essential for the user to perform tasks on a mobile UI.

\subsubsection{Agent solicits information from user}
Mobile UIs often request users to input information relevant to their goals. For example, \textit{destination city} or \textit{travel dates} in the scenario of booking a hotel. On mobile UIs, the information request is typically made through input text fields. A conversational agent should be able to similarly solicit essential information from users using natural language. For example, asking users questions like \textit{"Which hotel do you want to search for."} or \textit{"What is the check-in date of your stay?"} We refer to this type of task as \textit{Screen Question-Generation} since the questions should be generated based on the UI screen contexts. 

\subsubsection{Agent provides information to user}
A key function of GUIs is to convey information to users through visual means. Similarly, conversational agents should be able to articulate the information presented in GUIs using language. Given that UIs contain a vast amount of information and user needs vary \cite{todi21conversations}, there are numerous ways to deliver screen information. An example is \textit{Screen Summarization} \cite{10.1145/3472749.3474765}, which provides a short description of the purpose of the current screen, e.g., \textit{"A list of hotel rooms available at W San Francisco."}, or \textit{"A step-by-step recipes of butter chicken"}. The descriptions can help users quickly understand the UI when visual information is unavailable.

\subsection{User-initiated Conversation}
Users can also initiate conversations to request information or proactively provide information for the agent to process. 
\subsubsection{User solicits information from the agent}
 Users should be able to solicit screen information from the agent through conversation. When the information request is done through questions such as \textit{"What's the rate of the hotel room with a king-size bed?"}, the agent should respond with appropriate answers such as \textit{"\$330 per night"}, based on the information presented on the screen. We call this type of conversational interaction \textit{Screen Question-Answering}, similar to visual \cite{antol2015vqa} or text question-answering \cite{rajpurkar2016squad} but instead based on a mobile UI. \textit{Screen Question-Answering} is beneficial in situations where there is a large amount of text on a screen, such as search results, making it difficult to locate specific information. It is also useful for individuals with disabilities who rely on screen readers to access screen text. Instead of waiting for the screen reader to scan through all the content on the screen until the relevant information is found, they can simply ask for the needed information.

\label{user_solicits_info}
\subsubsection{User provides information to the agent}
Users can initiate conversations with the agent by providing new information. After receiving the messages, the agent should respond accordingly based on the current UI context using language and/or mobile actions. A representative task of this type of conversation is \textit{Mapping Instruction to UI Action.} For example, when a user who is presented with a hotel booking screen says, \textit{"Click on the Reserve button to book the Fabulous King room"}, the agent should understand the user's intent and the UI contexts in order to click the corresponding button. This type of interaction allows users to control the devices when touch inputs are unavailable \cite{wobbrock2019situationally}.
\label{user_provide_info}

\begin{figure*}[ht]
  \centering
  \includegraphics[width=1\linewidth]{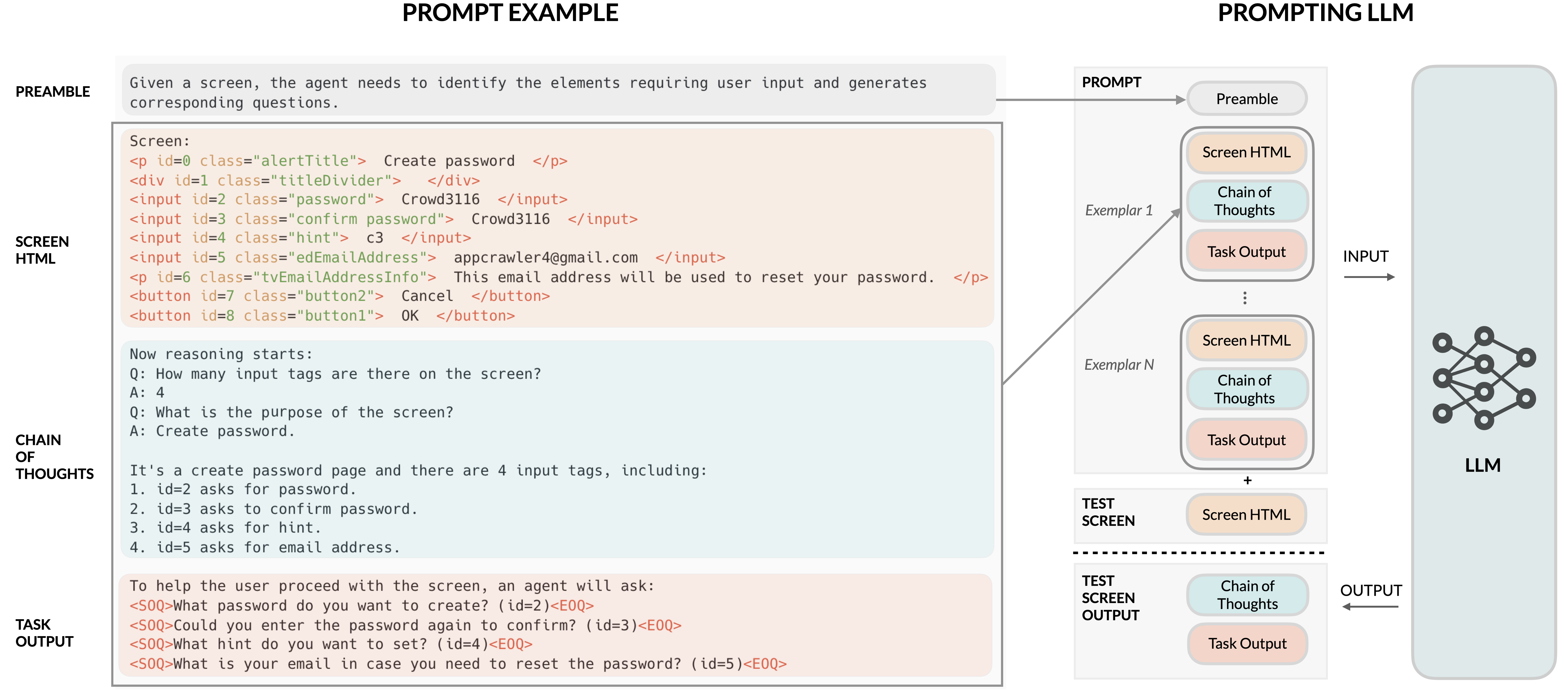}
  \caption{Left: An example illustrating the proposed prompt structure. A prompt starts with the preamble, which describes the task. Following the preamble, there will be zero or more task exemplars from the target tasks. Each exemplar consists of an input screen HTML, a chain of thoughts (if applicable), and a task-specific output. Additional exemplars are appended to the end of the previous ones. Right: An illustration of prompting large language models in our use cases. The prompt contains N exemplars from the target tasks. We append the screen HTML of the test screen to the end of the prompt. We then feed the prompt + test screen HTML as input into the LLM. The LLM then generates word tokens in an auto-regressive manner, which helps them capture the exemplars' pattern to produce a chain of thoughts (if applicable) and task-specific output.}
  \Description{The left shows an example illustrating the proposed prompt structure. A prompt starts with the preamble, which describes the task. Following the preamble, there will be zero or more task exemplars from the target tasks. Each exemplar consists of an input screen HTML, a chain of thoughts (if applicable), and a task-specific output. Additional exemplars are appended to the end of the previous ones. The right shows An illustration of prompting large language models in our use cases. The prompt contains N exemplars from the target tasks. We append the screen HTML of the test screen to the end of the prompt. We then feed the prompt + test screen HTML as input into the LLM. The LLM then generates word tokens in an auto-regressive manner, which helps them capture the exemplars’ pattern to produce task-specific output.}
  \label{prompt}
\end{figure*}

\section{Prompting Large-Language Models for Mobile UI Tasks}
\label{sec:prompting_techniques}
 We introduce a class of prompting techniques designed to adapt LLMs to mobile UIs to enable conversational interaction. LLMs support in-context few-shot learning via \textit{prompting}---instead of finetuning or re-training models for each new task, one can prompt an LLM with a few input and output data exemplars from the target task \cite{brown2020language, chowdhery2022palm, wei2022chainofthought, zhou2022leasttomost}. For many natural language processing tasks such as question-answering or translation, few-shot prompting performs competitively with benchmark approaches \cite{brown2020language}. However, the methodology for prompting LLMs with mobile UIs has yet to be established and presents several challenges. Firstly, language models can only take text input, while mobile UIs are multimodal, containing text, image, and structural information in their view hierarchy data and screenshots. \bw {Moreover, directly inputting the view hierarchy data of a mobile screen into LLMs is not feasible as it contains excessive information, such as detailed properties of each UI element, which can easily exceed the input length limits of LLMs.} Furthermore, the mobile UIs encapsulate the logic of target user tasks \cite{li20218kite}. Therefore, logical reasoning based on UI contexts is essential for the model to support conversations toward task completion. These unique aspects of mobile UIs pose two open problems for designing prompts:
 \begin{enumerate}
     \item How to represent mobile UIs in texts to leverage the few-shot prompting capability of LLMs?
     \item How to elicit reasoning based on the mobile UIs when needed?
 \end{enumerate}
We respond to these questions by describing our proposed prompting techniques and their design rationales. Prompting LLMs remains an ongoing research problem. As the first to investigate prompting with mobile UI, we provide a strong baseline approach and encourage future work to build upon our design and further study the open problems.

\subsection{Screen Representation}
\subsubsection{Representing View Hierarchy as HTML}
There are various ways to represent a mobile UI in text, e.g., concatenating all the text elements on the UI into a token sequence or using natural language sentences to describe UI elements, such as \textit{"a menu button in the top left corner."} To design our screen representation, we leverage the insight that if a prompt falls within the training data distribution of a language model, it is more likely for few-shot learning to perform. This is because LLMs are trained to predict the subsequent tokens that maximize the probability based on the training data. LLMs' training data is typically scraped from the web, including both natural language and code. For example, 5\% of PaLM's \cite{chowdhery2022palm} training data was scraped from GitHub, including 24 common programming languages such as Java, HTML, and Python \cite{chowdhery2022palm}. Therefore, we represent a mobile UI in texts by converting its view hierarchy data into HTML syntax. HTML is particularly suitable for representing mobile UIs as it is already a markup language representing web UIs. The conversion is conducted by traversing the view hierarchy tree using a depth-first search. We detail our conversion algorithm in the following sections. Note that since the view hierarchy is not designed to be represented in HTML syntax, a perfect one-to-one conversion does not exist. In contrast, our goal is to make the converted view hierarchy similar to the HTML syntax to generate a data representation closer to the training data distribution. 

\subsubsection{View Hierarchy Properties}
Converting a mobile UI's view hierarchy into HTML syntax can preserve the detailed properties of UI elements and their structural relationship. The view hierarchy is a structural tree representation of the UI where each node, corresponding to a UI element, contains various properties such as the class, visibility-to-user, and the element's bounds. However, using all element properties will result in lengthy HTML text, which may exceed the input length limit of the language model, e.g., 1920 tokens for PaLM and 2048 tokens for GPT-3. Therefore, we use a subset of properties related to the text description of an element:

\begin{itemize}
 \item \texttt{class}: Android object type such as \texttt{TextView} or \texttt{Button}.
 \item \texttt{text}:  element text that is visible to the user. 
 \item \texttt{resource\_id}: text identifiers that describe the referenced resource.
 \item \texttt{content\_desc}: content description that describes the element for accessibility purposes, i.e., the alt-text.
\end{itemize}
\subsubsection{Class Mapping}
We developed heuristics to map the Android classes to HTML tags with similar functionalities. We map \texttt{TextView} to the \texttt{<p>} tag as they are both used for presenting texts; all button-related classes such as \texttt{Button} or \texttt{ImageButton} are mapped to \texttt{<button>}. We map all image-related classes such as \texttt{IMAGEVIEW} to \texttt{<img>}, including icons and images.  Lastly, we convert the text input class \texttt{EDITTEXT} to \texttt{<input>} tag. We focus on the most common element classes for simplicity, and the rest of the Android classes, including containers such as \texttt{LinearLayout} are mapped to the \texttt{<div>} tag.  

\subsubsection{Text, Resource\_Id, and Content Description}
We insert the \texttt{text} properties of Android elements in between the opening and closing HTML tags, following the standard syntax of texts in HTML. The \texttt{resource\_id} property contains three entities: {package\_name}, {resource\_type}, and {resource\_name}. Among them, {resource\_name} usually contains additional descriptions of an element's functionality or purpose written by the developers. For example, in the Gmail app, an element with {resource\_name} of \texttt{"unread\_count\_textView"} shows how many emails are unread, whereas a \texttt{"date"} means the element shows the date of receiving a mail. Such information helps the model to better understand the screen context. We insert the resource\_name tokens that describe each element's purpose as additional identifiers in the \texttt{"class"} attributes, which originally contain identifiers linked to a style sheet or used by JavaScript to access the element. Word tokens in resource\_name are typically concatenated with underscores, which we replace as spaces when inserting. Lastly, we insert the \texttt{content\_desc} as the \texttt{"alt"} attribute in the HTML tags when the property is present.

\subsubsection{Numeric Indexes for Referencing}
To help model referencing specific UI elements, we insert numeric indexes to each element as the \texttt{"id"} attribute. The indexes are generated with the depth-first search order in the view hierarchy tree. For tasks such as predicting which button to click based on language instructions, the model can refer to elements using numeric indexes, which is more efficient and space-saving than spelling out the complete HTML tag.

\subsection{Chain-of-Thought Prompting}
Mobile UIs encapsulate the logic of user tasks \cite{li20218kite}; therefore, it is vital for models to perform reasoning when used for conversational interaction. LLMs have demonstrated abilities to reason \cite{brown2020language, chowdhery2022palm, 2022saycan} as they captured real-world knowledge during training with a large number of texts. Recent work further shows that LLM's reasoning ability can be improved by generating and chaining intermediate results to obtain the final answers, namely, Chain-of-Thought prompting \cite{wei2022chainofthought}. The idea is straightforward, i.e., simply appending a chain of thoughts describing intermediate results before the answers in the prompt. The model would then follow the patterns to generate a chain of thoughts during inference. Chain-of-Thought prompting has been shown to be helpful for reasoning tasks. The results are also more interpretable as the model would articulate its thought process before coming up with the answer. However, prior work has not investigated whether it can facilitate reasoning in generating conversations based on mobile UIs. Therefore, we incorporate the method in our experiments. \bw {Chain-of-Thought prompting has been shown  effective for tasks that require the model to reason across multiple steps. In our early investigations, we found that this technique does not improve performance for tasks where the output can be directly obtained from the input screen HTML. Therefore, in our experiments, we only used the technique for  \textit{Screen Question-Generation} whose task setup requires multi-step reasoning.}

\subsection{Prompt Structure}
We follow a similar prompt structure proposed in \cite{brown2020language}. Each prompt starts with a \textit{preamble} which explains the prompt's purpose. The preamble is followed by multiple exemplars consisting of the input, a chain of thought (if applicable), and the output for each task. Each exemplar's input is a mobile screen in the HTML syntax. To better leverage few-shot learning while complying with LLM's input length limits, we only show the leaf nodes visible to the users, as non-leaf nodes are usually containers that do not contain textual information. Following the input, a chain of thoughts is provided to elicit logical reasoning from LLMs, if applicable to the task. The output is the desired outcome for the target tasks, e.g., a screen summary or an answer to the question asked by the user. Figure \ref{prompt}-left shows an example of a 1-shot prompt. Few-shot prompting can be achieved with more than one exemplar included in the prompt. During prediction, we feed the model with the prompt with a new input screen appended at the end. Therefore, for $N$-shot learning, the prompt will consist of a preamble, $N$ exemplars, and the test screen for prediction, as shown in Figure \ref{prompt}-right.

\section{Feasibility Experiments}
\begin{figure}[ht]
  \centering
  \includegraphics[width=1\linewidth]{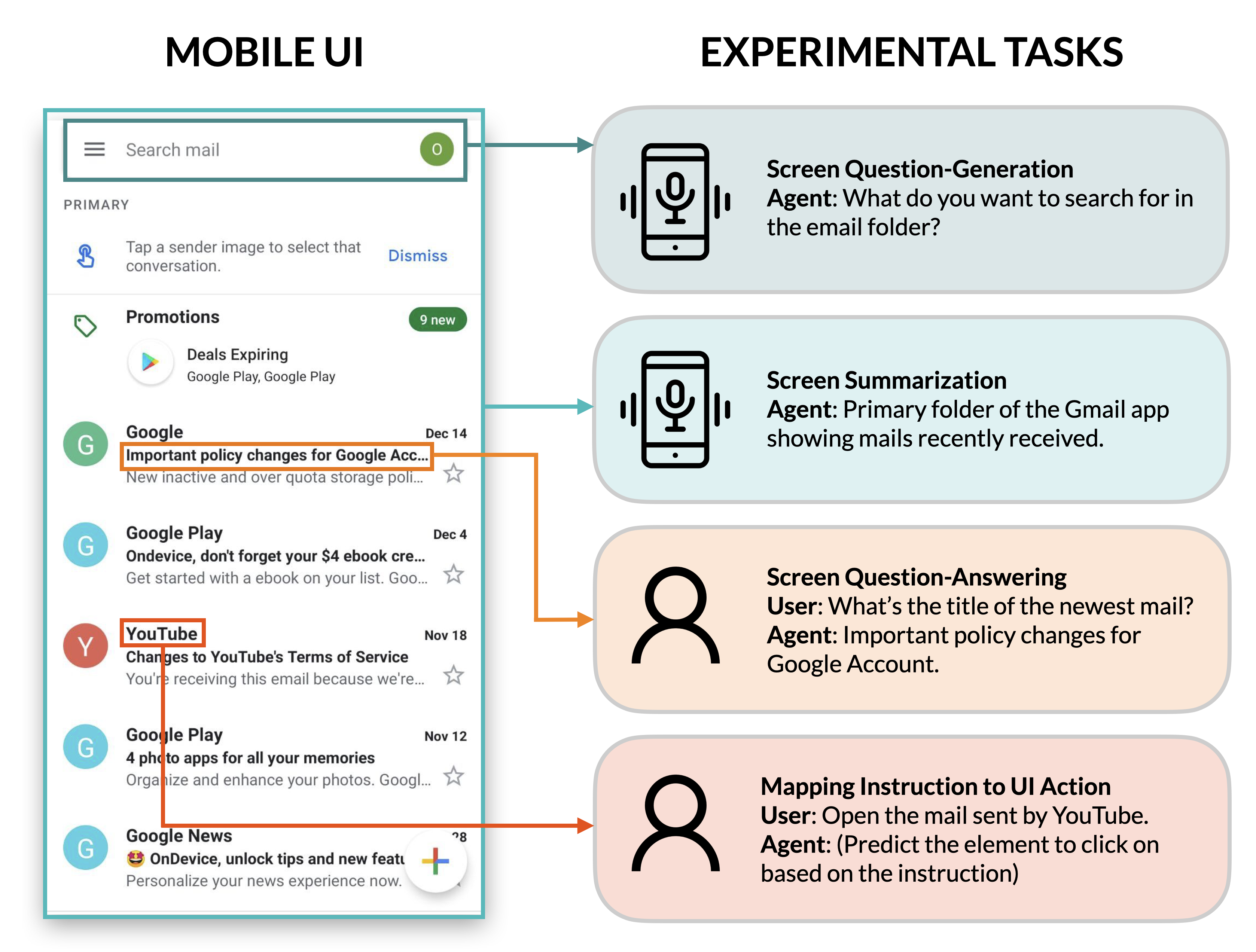}
  \caption{An illustration of the four UI modeling tasks we experimented on a single mobile UI. The tasks include \textit{Screen Question-Generation}, \textit{Screen Summarization}, \textit{Screen Question-Answering}, and \textit{Mapping Instruction to UI Action}. Each task is associated with a conversation category described in Section \ref{conversation_space}. Bounding boxes on the mobile UI highlight elements relevant to the example conversational interactions from each task. }
  \Description{This figure shows a Gmail UI screen with 4 bounding boxes on it. Each bounding box is associated with an experimental task tested in the feasibility study. The search bar is associated with the Screen Question-Generation task, as the agent should ask, “What do you want to search for in the email folder?”. The full screen is associated with the screen summary as the agent should summarize it as “Primary folder of the Gmail app showing mails recently received”. The latest email’s title is associated with the Screen Question-Answering as the user asks, “what’s the title of the newest mail?” and the agent should respond with “Important policy changes for Google Account”. A mail sender is associated with the Mapping instruction to UI Action task as the user commands “Open the mail sent by YouTube,” The agent should predict the element to click on it.}
  \label{tasks}
\end{figure}

As shown in Figure \ref{tasks}, we demonstrate the feasibility of using LLMs to enable conversations on GUIs through experiments with four tasks we introduced in section \ref{conversation_space}: 1) \textit{Screen Question-Generation}, 2) \textit{Screen Summarization}, 3) \textit{Screen Question-Answering}, and 4) \textit{Mapping Instruction to UI Action}.  Following the common practices of few-shot prompting \cite{wei2022chainofthought, brown2020language}, we select a handful of exemplar data to construct prompts for each task. We then evaluate the effectiveness using task-specific metrics detailed in each experiment. All the studies were conducted with the PaLM model \cite{chowdhery2022palm}, which performs similarly to other LLMs such as GPT-3 \cite{wei2022chainofthought}. The PaLM model is trained with a maximum input length of 1920 tokens. Therefore, we limit the number of exemplars in a prompt to be \bw{two at most}, excluding the test screen, to avoid exceeding the length limit. The experiments aim to understand what can be achieved by simply prompting LLMs with a few exemplars from the target tasks and compared to the baseline or benchmark if available.

\label{exp}

\begin{figure*}[ht]
  \centering
  \includegraphics[width=1\linewidth]{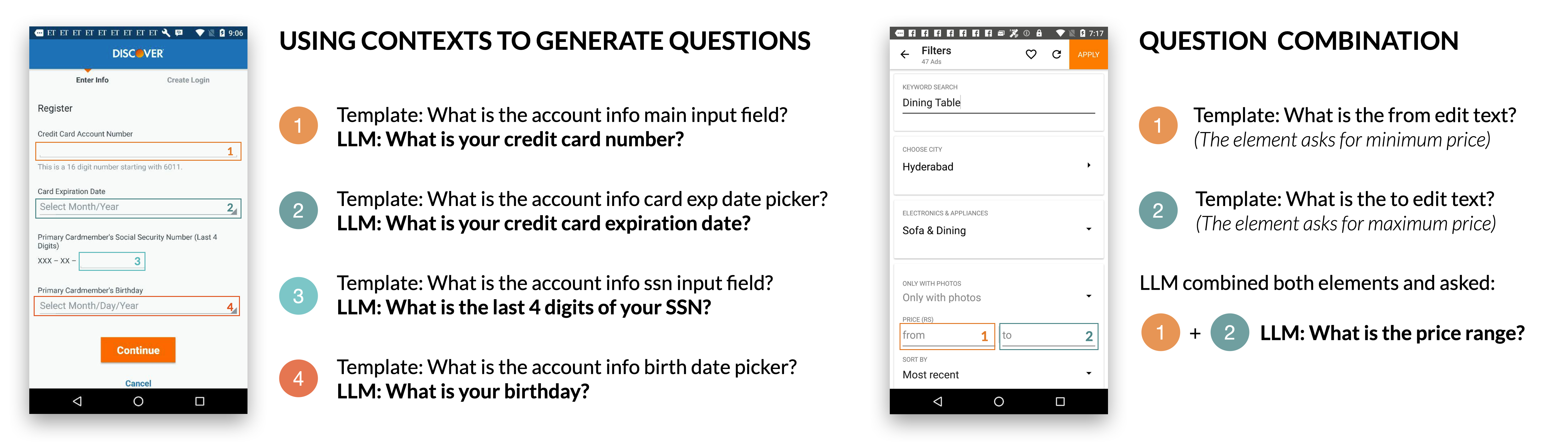}
  \caption{Example screen questions generated by the LLM. Left: The LLM can utilize screen contexts to generate grammatically-correct questions relevant to each input field on the mobile UI, while the template approach falls short. Right: We observed that the LLM could use its prior knowledge to combine multiple related input fields to ask a single question. The example shows the LLM combining the minimum and maximum prices fields into a single question asking about the price range. Elements relevant to each question are highlighted in corresponding colors and numbered indexes.}
  \Description{Example screen questions generated by the LLM. Left: The LLM can utilize screen contexts to generate grammatically-correct questions relevant to each input field on the mobile UI, while the template approach falls short. For example, for a text field asking about credit card numbers, the LLM generates a question asking, “What’s your credit card number while the Template approach asks, “What is the account info main input field?” Right: We observed that the LLM could use its prior knowledge to combine multiple related input fields to ask a single question. The example shows the LLM combining the minimum and maximum prices fields into a single question asking about the price range.}
  \label{q}
\end{figure*}

\subsection{Screen Question-Generation}
\subsubsection{Task Formulation} 
Given a mobile UI screen, the goal of screen question-generation is to synthesize coherent, grammatically-correct natural language questions relevant to the UI elements requiring user input. The task occurs when the agent requests user input to proceed on the UI.

\subsubsection{Prompt Construction}
Figure \ref{prompt} shows an example prompt we used to generate questions. We use a preamble of \textit{"Given a screen, the agent needs to identify the elements requiring user input and generates corresponding questions."}. We used chain-of-thought techniques to generate three intermediate results 1) input fields count, 2) screen summary, and 3) input enumeration. The input field counts were fed with a ground truth count extracted from the screen HTML. We found this step essential to prevent the model from omitting some input elements and only generating a subset of questions. Next, we asked the model to summarize the screen's purpose, which produces the screen context that helps provide details in the generated questions. Subsequently, the model enumerates which elements are asking for what information. After the chain of thoughts, the model generates the questions, enclosed by \texttt{<SOQ>} and \texttt{<EOQ>} tokens, representing start-of-question and end-of-question, respectively. The tokens are used as delimiters for conveniently parsing the questions from the output texts generated by the model. We use similar ways to insert special tokens for parsing model output in the rest of the experiments. An example prompt can be found in appendix \ref{prompt:1}.

\subsubsection{Experimental Setup}
\label{sqp-exp}
We aim to understand the quality of LLMs for natural language generation based on UI elements. Since there is currently no existing dataset for screen question generation, we followed the common practice of evaluating language generation quality with human ratings. We randomly sampled 400 test screens from the RICO dataset \cite{Deka:2017:Rico}. Each of these screens contains at least one \texttt{EditText} element, representing the text input field for users to enter information on the UI. We randomly selected another two screens from the RICO dataset as exemplars to include in the prompt. An \texttt{EditText} element represents an input field for the user to enter information, and we generate questions for every input field. We generate questions from the test screens using a prompt constructed with two exemplars. Some screens contain multiple input fields, and sometimes several of them are relevant and can be asked collectively. For example, three fields asking for the birth year, month, and date can be combined into a single question as \textit{"when is your birthday?"}. Combining questions can lead to a more efficient conversation between an agent and a user. Therefore, we include an exemplar that combines relevant questions in the prompt to see if the model will also learn to combine relevant questions.

We compare LLM's results with a rule-based approach that uses words in \texttt{resource\_id}, referred to as \textit{res\_tokens}, to fill in the template of \textit{"What is \{res\_tokens\}?"}. We use \textit{res\_token} instead of \textit{text} because most text input fields are blank by default, and \textit{res\_token} contains the most meaningful description of an input field.  We recruited 17 raters who work as professional data labelers at Google to provide ratings. To ensure the quality of the labels, a group of quality audits sampled and reviewed ~5\% of the total number of questions answered by every rater. We provided a UI screenshot of a mobile UI and a generated question for each labeling task. The \texttt{EditText} element associated with the question is highlighted with a bounding box. We solicited human ratings on whether the questions were grammatically correct and relevant to the input fields for which they were generated. In addition to the human-labeled language quality, we automatically examined how well LLMs can cover all the elements that need to generate questions. The evaluation metrics include the following:

\begin{itemize}
     \item \textbf{Grammar Correctness}: How correct is the grammar of a generated question? Are the sentences intelligible and plausible? This metric tests the language generation quality in general and is rated on a 5-point Likert scale with 1 as completely incorrect and 5 as completely correct. 
    \item \textbf{UI Relevance}: Whether a generated question is relevant to the highlighted UI element. This metric tests whether the connection between a UI element and a question is correctly established by the model, which is rated on a binary scale as either relevant or not relevant. 
    \item \textbf{Question Coverage}: How well can the model identify the elements on the screen that need question generation? This metric is automatically computed by comparing the indices of ground truth input elements with those identified by the model within the chain of thoughts. 

\end{itemize}

\begin{table}
  \caption{Grammar correctness, UI relevance, and question coverage results from the screen question-generation experiment.}
  \label{tab:human}
  \begin{tabular}{lccc}
    \toprule
    Method& Grammar & Relevance & Coverage F1\\
    \midrule
    Template & 3.60 ($\sigma$=0.69) & 84.1\% & 100\%\\
    LLM & 4.98 ($\sigma$=0.07) & 92.8\% & 95.9\%\\
  \bottomrule
  \end{tabular}
\end{table}

\subsubsection{Results}
We evaluated 931 questions for both the LLMs and the template-based approach. Three different human raters examined each question to obtain aggregated scores. Table \ref{tab:human} shows the results of our evaluation. Our approach achieves an almost perfect average score of 4.98 on grammar correctness, while the rule-based approach receives a 3.6 average rating. A Mann–Whitney U test shows that the difference between the two methods is statistically significant (p < 0.0001). LLMs also generate 8.7\% more relevant questions compared to the baseline. In terms of questions coverage, our approach achieves an F1 score of 95.9\% (precision = 95.4\%, recall = 96.3\%). Since the rule-based method iterates through every input field to generate questions, its question coverage is naturally 100\%. Altogether, the results show that our approach can precisely identify input elements and generate relevant questions that are intelligible. We further analyzed the model behaviors, and our results revealed interesting emergent abilities of LLMs. When generating a question for a field, the model considers both the input field element and the \textit{screen context} (information from other screen objects). For example, Figure \ref{q} shows how the model leveraged screen contexts to generate four questions for the input fields on a credit card register screen. While the baseline outputs use the \textit{ref\_tokens} to convey somewhat relevant information, they are less intelligible than the LLM output and do not articulate the specific information requested by the fields. In contrast, all four questions generated by LLM are grammatically correct and ask for relevant information.  For Question 3 in Figure \ref{q}, the LLM additionally uses the texts above the input field to ask for the \textit{"last 4 digits of SSN"}. The model also blends the screen contexts into the generated questions. For instance, Question 2 asks for "credit card expiration date," while the texts above did not mention the word "credit." We also observed that the model exhibits the behavior of combining relevant fields into a single question on three test screens. For example, Figure \ref{q} shows the model can combine two input elements asking for minimum and maximum values for price into a single question \textit{"What is the price range?"}. In practice, combining questions can lead to more efficient communication between users and agents.

\begin{table*}
  \caption{Screen summarization performance on automatic metrics. }
  \label{tab:sum_number}
  \begin{tabular}{lcccccccc}
    \toprule
        Model & BLEU-1 & BLEU-2 & BLEU-3 & BLEU-4 & CIDEr & ROUGE-L &  METEOR  \\
    \midrule
        0-shot LLM & 7.8 & 6.4 & 5.9 & 5.7 & 1.5 & 3.4 & 4.5 \\ 
        1-shot LLM & 42.3 & 21.1 & 14.8  & 12.1 & 40.9 & 28.9 & 15.3  \\ 
        2-shot LLM & 45.0 & 25.1 & 17.6 & 14.1 & 39.9 & 33.0 & 17.7 \\ 

        Screen2Words~\cite{10.1145/3472749.3474765} & 65.5 & 45.8 & 32.4 & 25.1 & 61.3 & 48.6 & 29.5  \\ 
  \bottomrule
    \end{tabular}
\end{table*}

\subsection{Screen Summarization}
\label{summarization}
\subsubsection{Task Formulation}
Screen summarization was proposed in \cite{10.1145/3472749.3474765} as the automatic generation of descriptive language overviews
that cover essential functionalities of mobile screens. The task helps users quickly understand the purpose of a mobile UI, which is particularly useful when the UI is not visually accessible. 

\subsubsection{Prompt Construction} We use a preamble of \textit{"Given a screen, summarize its purpose."} We did not use chain-of-thought prompting as no intermediate result is needed to be generated for the task. We used $N$ pairs of screen HTML and corresponding summary following the preamble, where $N=0,1,2$ represents $N$-shot learning. The output summaries are enclosed by special tags <SOS> and <EOS>, meaning the start and end of a summary, respectively. An example prompt can be found in appendix \ref{prompt:2}.

\subsubsection{Experiment Setup}
We use the Screen2Words dataset \cite{10.1145/3472749.3474765} to test LLM's ability to summarize screens. The dataset contains human-labeled summaries for more than 24k mobile UI screens, each with five summary labels. To gauge the quality of our approach, we test the performance of using LLMs to summarize screens with Screen2Words' test set, consisting of 4310 screens from 1254 unique apps, which was used by the benchmark. We randomly sampled two screens from the dataset and one of their corresponding summaries as the exemplars for prompt construction. We use the same automatic metrics reported in the original paper, including BLEU, CIDEr, ROUGE-L, and METEOR. \bw{As prior work \cite{10.1145/3472749.3474765, gptsummarization} has found that automatic scores may not correlate well with human perception of summary quality, we additionally conducted a human evaluation to solicit subjective feedback on the summaries generated by both LLM and the benchmark model Screen2Words \cite{10.1145/3472749.3474765}. We recruited 37 annotators using the same process specified in section \ref{sqp-exp}. We presented annotators with a mobile UI screen and two screen summaries during the labeling process. To avoid bias, we randomly assigned the summaries generated by LLM and Screen2Words to be either Summary 1 or Summary 2, without revealing which summary was generated by which model. We instructed annotators to choose the summary that best \textit{accurately} summarizes the mobile screen. The annotators were given three options: (a) Summary 1, (b) Summary 2, and (c) Equal or Very Similar. They were instructed to choose the third option only when it was difficult to judge the differences between the quality of the two summaries. The rating study was conducted on the full test dataset, with each screen receiving three ratings from different annotators.}

\begin{figure}[ht]
  \centering
  \includegraphics[width=1\linewidth]{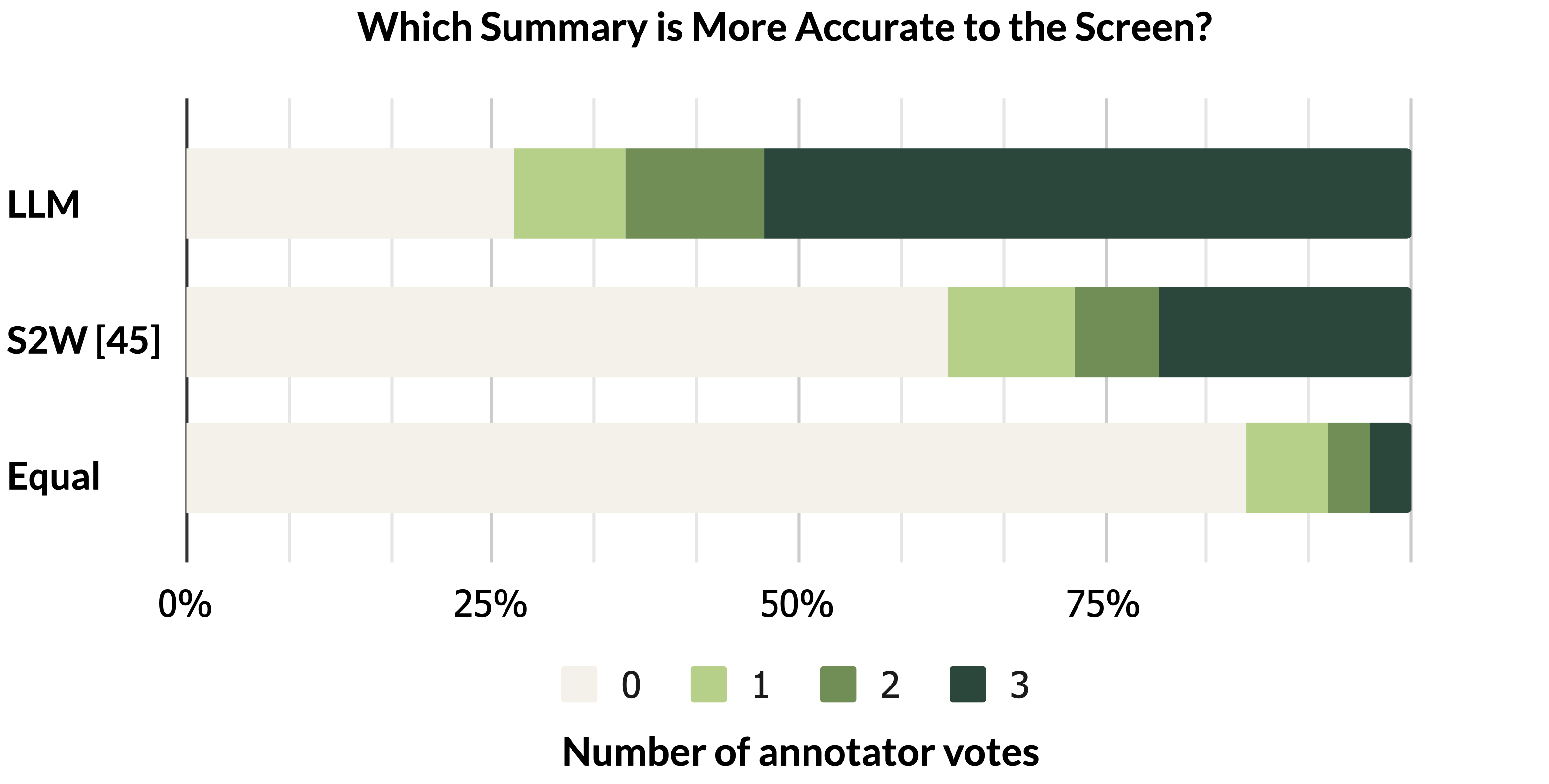}
  \caption{Annotator vote distribution across all test screens. LLM summaries are more accurate than those of the benchmark model for 64.1\% of screens across the test set. This choice is unanimous between three labelers for more than half of the screens (52.8 \%).}
  \Description{Annotator vote distribution across all test screens. LLM summaries are more accurate than the benchmark model for 64.1\% of screen across the test set. This choice is unanimous between three labelers for more than half of the screens (52.8 \%).}
  \label{summary_rating}
\end{figure}

\begin{figure*}[ht]
  \centering
  \includegraphics[width=1\linewidth]{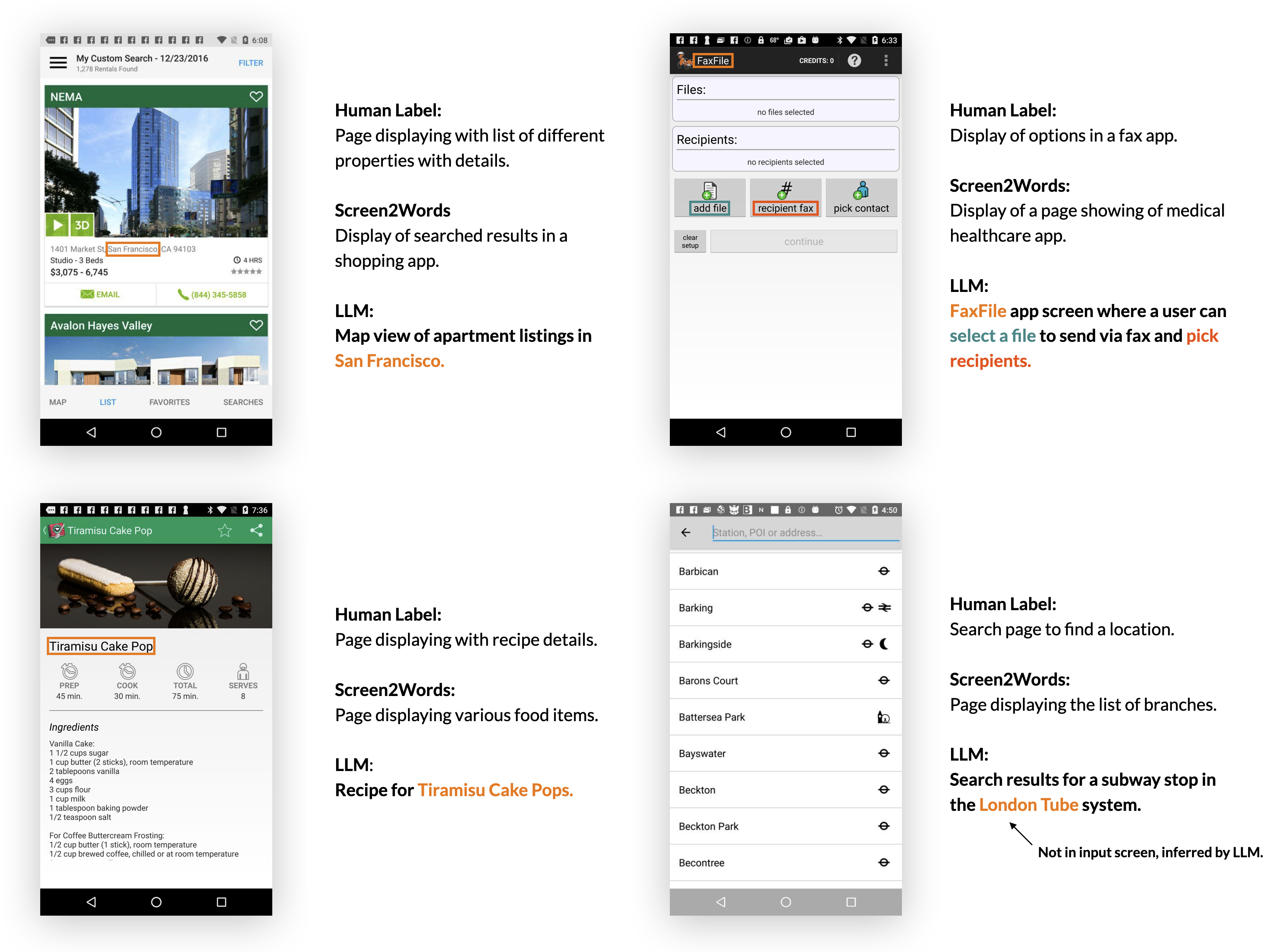}
  \caption{Example summaries generated by prompting the LLM with two exemplars (2-shot learning). The LLM is likelier to use specific texts on the screen to compose summaries (top-left and bottom-right). Moreover, the LLM is more likely to generate more extended summaries that leverage multiple vital elements on the screen (top-right). We also observed that the LLM would use its prior knowledge to help summarize the screens. For example, the bottom-right shows the LLM had inferred the screen is for the London Tube system from the station names displayed on the screen. UI elements relevant to the highlighted phrases in the summaries are called out by bounding boxes with corresponding colors.}
  \Description{Example summaries generated by prompting the LLM with 2 exemplars (2-shot learning). The LLM is likelier to use specific texts on the screen to compose summaries (top-left and bottom-right). Moreover, the LLM is more likely to generate more extended summaries that leverage multiple vital elements on the screen (top-right). We also observed that the LLM would use its prior knowledge to help summarize the screens. For example, the bottom-right shows the LLM had inferred the screen is for the London Tube system from the station names displayed on the screen.}
  \label{summary}
\end{figure*}

\subsubsection{Results}
Table \ref{tab:sum_number} shows the results of screen summarization on automatic metrics. The model could not generate meaningful summaries in the zero-shot setting (not providing any exemplar). This result is expected as the LLM's training data may not have covered the task of screen summarization. When provided with the model one exemplar, the performance significantly boosted across all metrics. More examples in the prompt provide marginally higher scores. 
\bw{The average length of summaries generated by our two-shot LLM model was 7.15 words (STD=2.68). In comparison, the average length of summaries generated by the benchmark model was 6.64 (STD=1.98). Additionally, the LLM used a significantly larger number of unique words, at 3062, compared to the 645 unique words used by the benchmark model. The results indicate that LLM is capable of generating longer summaries with a wider range of language. When evaluating these models against automatic metrics, LLM generally scored less than the benchmark model. This is expected because the benchmark model was trained with the Screen2Words dataset, and the automatic metrics are based on token matching. A model trained on the dataset could achieve high scores by prioritizing the high-frequency words in the dataset. However,  many high-frequency words and phrases, such as "display of," "screen', and "app", in the Screen2Words dataset do not meaningfully contribute to the summarization accuracy. As a result, a specialized model that learned to prioritize these frequent words may score well by synthesizing generic summaries, indicating the limitations of existing evaluation metrics.}

In contrast, our human evaluation revealed that LLM generates summaries with higher perceived quality than those generated by the benchmark model, Screen2Words. Among all the human annotations, 63.5\% of them rated LLM summaries as more accurate, while 28.6\% voted for benchmark model summaries. Moreover, 7.9\% of annotations indicated that LLM and Screen2Words generate summaries with equal or very similar accuracy. We further examine the annotator agreement across all individual screens. As shown in Figure \ref{summary_rating}, LLM summaries are deemed as more accurate for 64.1\% of screens according to the majority vote (> 2 votes). Remarkably, LLM summaries were rated as more accurate unanimously by three annotators for 52.8 \% of screens. However, our study also showed the LLMs summaries are not always better than the benchmark model, whose summaries received unanimous votes for 20.5\% of screens. The results suggest that improving our approach may involve incorporating visual information \cite{flamingo} (as Screen2Words does), as current language models do not have access to this type of information. Therefore, our approach may not perform well for screens without texts and have a strong focus on visuals.

\begin{table*}
  \caption{The LLM's performance on the screen QA task.}
  \label{tab:qa_number}
  \begin{tabular}{lcccc}
    \toprule
        Model & Exact Matches & Contains GT & Sub-String of GT & Micro-F1 \\
    \midrule
        0-shot LLM & 30.7\% & 6.5\% & 5.6\% & 31.2\%\\
        1-shot LLM & 65.8\% & 10\% & 7.8\% & 62.9\% \\  
        2-shot LLM & 66.7\% & 12.6\% & 5.2\% & 64.8\% \\
        DistilBERT~\cite{Sanh2019distilbert} & 36.0\% & 8.5\% & 9.9\% & 37.2\% \\

  \bottomrule
    \end{tabular}
\end{table*}

Figure \ref{summary} shows example screens with summaries annotated by human labelers and the output from both Screen2Words and the LLM model. We found that the LLM is more likely to use specific texts on the screen to compose summaries, such as San Francisco (top-left) and Tiramisu Cake Pop (bottom-left), while the Screen2Words dataset and the benchmark model output tend to be more generic. Moreover, the LLM is more likely to generate more extended summaries that leverage multiple vital elements on the screen. For example, the top-right screen shows that the LLM composes a longer summary by leveraging the app name, the send file button, and the recipient fax button: \textit{"FaxFile app screen where a user can select a file to send via fax and pick recipients."} We also observed that the LLM's prior knowledge is beneficial for screen summarization. For example, the bottom-right photo shows a station search results page for London's tube system. The LLM predicts \textit{"Search results for a subway stop in the London tube system."} However, the input HTML does not contain the words \textit{"London"} nor \textit{"tube."} Therefore, the model has utilized its prior knowledge about the station names, learned from large language datasets, to infer that they belong to the London Tube. This type of summary may not have been generated if the model is trained only on the Screen2Words dataset and shows the benefit of leveraging LLM's prior knowledge for summarizing UIs.

\begin{figure*}[ht]
  \centering
  \includegraphics[width=1\linewidth]{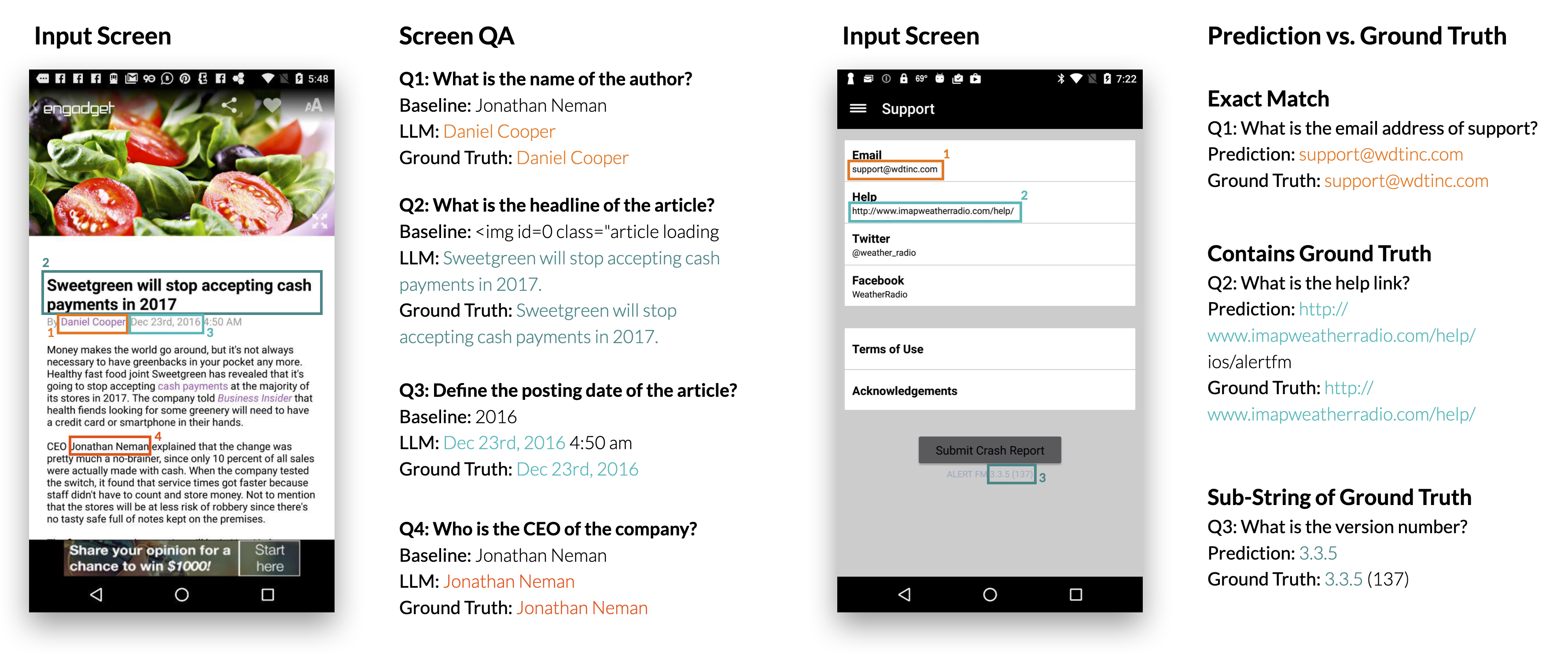}
  \caption{Left: Example results from the screen QA experiment. The LLM significantly outperforms the off-the-shelf QA model DistillBert (Baseline). Right: Example of the three metrics used to assess the LLM's performance on screen QA, including Exact Match, Contains Ground Truth, and Sub-String of Ground Truth. For each example question, the UI element related to its ground truth is highlighted with a bounding box with the corresponding color and question index. }
  \Description{Left: Example results from the screen QA experiment. The LLM significantly outperforms the off-the-shelf QA model DistillBert (Baseline). Right: Example of the three metrics used to assess the LLM’s performance on screen QA, including Exact Match, Contains Ground Truth, and Sub-String of Ground Truth. For each question, the UI element related to its ground truth is highlighted with a bounding box that has the corresponding color and question index.
}
  \label{fig:qa}
\end{figure*}

\subsection{Screen Question-Answering (QA)}
\subsubsection{Task Formulation}
Given a mobile UI and an open-ended question asking for information regarding the UI, the model should provide the correct answer. We focus on factual questions, which require answers based on facts presented on the screen.
\subsubsection{Prompt Construction} We use a preamble of \textit{"Given a screen and a question, provide the answer based on the screen information."}. We did not use chain-of-thought prompting as no intermediate result is needed to be generated for the task. In our experiments, $N$ sets of screen HTML and question-answer pairs follow the preamble, where $N=0, 1, 2$ represents $N$-shot learning. The output answers are enclosed by special tags <SOA> and <EOA>, meaning the start and end of an answer, respectively. The prompt we used in the study can be found in appendix \ref{prompt:3}.
\subsubsection{Experiment Setup}
We use a dataset of 300 human-labeled question-answer pairs from 121 unique screens in the RICO dataset \cite{Deka:2017:Rico}.
It is a preliminary dataset obtained from the authors of an on-going large-scale data collection for screen question-answering \cite{screenqa}. 
The data labeling process involves two stages: question annotation and answer annotation. For question annotation, the annotators were asked to frame questions given a screenshot as the context. The annotators were expected to compose questions that only inquire about information that requires no logical reasoning and can be directly read off from the screen. After that, another set of annotators answers the previously annotated questions given the associated screenshots. We randomly held out three screens for prompt construction. The three screens were associated with 12 QA pairs, and we randomly sampled one pair from each screen as exemplars to include in the prompt. In the remaining 288 question-answer pairs, 57 ground truth answers are not present in the view hierarchy data. This is expected because the labeling is based on screenshots instead of view hierarchies, and many screens in the RICO dataset contain inaccurate view hierarchy data \cite{li2022denoising}. In this work, we focus on the answers that are present in the view hierarchy data, and incorporating screenshot information in the LLM will be a critical direction to investigate in the future.

Since the answers are generated instead of retrieved from screen HTML, some correct answers may not completely match the labels. For example, \textit{"2.7.3"} versus \textit{"version 2.7.3"}. Therefore, we report performances on four metrics: 1) \textit{Exact Matches}: the predicted answer is identical to the ground truth. 2) \textit{Contains GT}: the answer is longer than the ground truth and fully contains it. 3) \textit{Sub-String of GT}: the answer is a sub-string of the ground truth. 4) \textit{Micro-F1}: the micro F1 scores, calculated based on the number of shared words between the predicted answer and the ground truth across the entire dataset. We consider \textit{Exact Match} answers correct, and those fall within \textit{Contains GT} and \textit{Sub-String of GT} relevant. The three metrics are exclusive to each other, and examples of each can be found in Figure \ref{fig:qa}-right.  We compare the LLM with an off-the-shelf text QA model DistilBERT \cite{Sanh2019distilbert}. We use the \texttt{distilbert-base-cased-distilled-squad} implementation on \texttt{huggingface.co}, which achieved 79.6\% Exact Match and 87.0\% F1 score on the SQuAD dataset \cite{rajpurkar2016squad}. Unlike existing QA models that extract answers from the input text, the LLM may generate answers in  aliases equivalent to the ground truth, for example, 9/15 and September 15th. However, our metrics, which rely on string matching, may not detect these aliases and could result in under-reported scores for our approach.

\subsubsection{Results}
Table \ref{tab:qa_number} shows the QA results of different settings. Unlike screen summarization, we found that the LLM can already perform screen QA with the zero-shot setting. 30.7\% of the generated answers match exactly with the ground truth, 6.5\% answers contain the ground truth, and 5.6\% answers are sub-strings of the ground truth. The zero-shot performance might be because the training data of LLMs already contain many QA-related data from the internet. Therefore, the model had already learned to perform question-answering. The off-the-shelf DistillBert model achieves 36\% Exact Match, 8.5\% Contains GT scores, and 9.9\% Contains GT scores, slightly better than the zero-shot performance of LLMs. DistillBert model performs much poorly on our tasks compared to standard question-answering benchmarks, which might be because it was not trained with HTML data. Similar to screen summarization, by providing a single exemplar, the performance boosted significantly, achieving 65.8\% Exact Match, 10\% Contains GT, and 7.8\% Sub-String of GT--summing up to 83.6\% answers relevant to the ground truth. 1-shot LLM achieved a 62.9\% Micro-F1 score, significantly outperforming the baseline's score of 37.2\%.
Once again, we observed that the 2-shot setting provided only a modest performance improvement, with relevant answers reaching 84.5\% and Micro-F1 achieving 64.8\%. Figure \ref{fig:qa}-left shows example QA results from our experiment using 2-shot learning. The LLM can effectively understand the screen and generate a correct or relevant answer. For the shown screen, the LLM correctly answers Q1, Q2, and Q4. For Q3, the LLM generates an answer containing the ground truth "Dec 23rd, 2016" but includes the time within the day "4:50" am. In contrast, the baseline model trained with standard text question-answering corpus only managed to answer Q4 correctly; for Q3, its answer only contains "2016", a sub-string of the ground truth. Q2 shows that the baseline model sometimes incorrectly retrieves HTML code from the input screen.

\subsection{Mapping Instruction to UI Action}
\label{study:4}
\subsubsection{Task Formulation}
Given a mobile UI screen and natural language instruction to control the UI, the model needs to predict the id of the object to perform the instructed action. For example, when instructed with "Open Gmail," the model should correctly identify the Gmail icon on the home screen. This task is useful for controlling mobile apps using language input such as voice access. 

\subsubsection{Prompt Construction} We use a preamble of \textit{ "Given a screen, an instruction, predict the id of the UI element to perform the instruction."}. The preamble is followed by $N$ exemplars consisting of the screen HTML, instruction, and ground truth id. Here $N=0, 1, 2$ representing $N$-shot learning. We did not use chain-of-thought prompting as no intermediate result is needed to be generated for the task. The output answers are enclosed by special tags <SOI> and <EOI>, meaning the start and end of the predicted element id, respectively. An example prompt can be found in appendix \ref{prompt:4}.

\subsubsection{Experiment Setup}
We use the PixelHelp dataset \cite{li-etal-2020-mapping}, which contains 187 multi-step instructions for performing everyday tasks on Google Pixel phones, such as switching Wi-Fi settings or checking emails. We randomly sampled one screen from each unique app package in the dataset as prompt modules. We then randomly sampled from the prompt modules to construct the final prompts. We conducted experiments under two conditions: 1) in-app and 2) cross-app. In the former, the prompt contains a prompt module from the same app package as the test screen, while in the latter, it does not. Following the original paper, we report the percentage of partial matches and complete matches of target element sequences.

\begin{table}
  \caption{Mapping Instruction to UI Action Results}
  \label{tab:mapping}
  \begin{tabular}{lcccc}
    \toprule
        Model & Partial & Complete \\
    \midrule
    0-shot LLM & 1.29 & 0.00  \\
    1-shot LLM (cross-app) & 74.69 & 31.67 \\
    2-shot LLM (cross-app) & 75.28 & 34.44 \\
    1-shot LLM (in-app)& 78.35 & 40.00  \\
    2-shot LLM (in-app)& 80.36 & 45.00 \\
    \midrule
        Seq2Act~\cite{li-etal-2020-mapping} & 89.21 & 70.59  \\
  \bottomrule
    \end{tabular}
\end{table}

\subsubsection{Results}
Our experimental results (Table \ref{tab:mapping}) show that the 0-shot setting cannot perform the task with nearly zero partial or complete accuracy. In the cross-app condition, one-shot prompting significantly achieves 74.69 partial and 31.67 complete, meaning ~75\% of elements associated with the instructions were correctly predicted, and more than 30\% tasks are entirely correct. The 2-shot setting offers incremental boosts for both metrics. In the in-app condition, both the 1-shot and 2-shot settings achieve higher scores than their counterparts in the cross-app condition. Our best-performing setting is the 2-shot LLM \& in-app, which achieves 80.36 partial and 45.0 complete accuracy scores. Our approach underperforms the benchmark results from the Seq2Act model \cite{li-etal-2020-mapping}, which was trained on several dedicated datasets with hundreds of thousands of examples. We do not expect prompting LLMs to consistently outperform benchmarks across all tasks, as the model can only view a handful of data examples and does not update its underlying parameters \cite{brown2020language}. Nevertheless, our approach still demonstrated competitive performances for the task using only two examples.

\section{Discussions and Future Work}
In this section, we discuss the implications of our investigation, the limitations of our approach, and how future work can build upon our work. 

\subsection{Implications for Language-Based Interaction}
 An important takeaway from our studies is that \textit{prototyping novel language interactions on mobile UIs can be as easy as designing a data exemplar}. As a result, an interaction designer can rapidly create functioning mock-ups to test new ideas with end users. Moreover, developers and researchers can explore different possibilities of a target task before investing significant efforts into developing new datasets and models. For example, as discussed in Section \ref{summarization}, the summaries in the Screen2Words dataset follow a particular sentence structure, while there are many other ways to summarize a screen. Our approach can allow researchers to quickly inspect various types of summaries and how they work in different scenarios before spending efforts developing dedicated datasets and models.  
 
 The conversational interaction enabled by our approach is beneficial for accessibility. It can also be blended with other input/output modalities, such as touch input and screen readers, to offer new possibilities for multi-modal interaction. Another use case of our approach is \textit{end-user prompting}. When a conversational agent fails to understand or carry out the tasks associated with the user's commands, prior work has leveraged programming by demonstration technique that asks users to provide tasks demonstrations to teach the agent \cite{li2017sugilite, li2019pumice}. \bw{Our work could augment these techniques to allow users to teach the model by prompting. With the assistance of our algorithm, users only need to provide examples of the desired output, such as which button to click given a language command. For more complex prompts, such as ones using Chain-of-Thought prompting, future work can explore using LLMs to assist users in prompt writing.}

\subsection{Feasibility Experiments}
\bw{Our approach showed promising results on each task in the feasibility experiments. Notably, our work contributes to the literature as the first to investigate methods to enable \textit{Screen Question-Generation} and \textit{Screen Question-Answering}. Moreover, our approach can generate screen summaries that are more accurate than the benchmark model, according to the human evaluation of the \textit{Screen Summarization} task. Although our method underperformed the benchmark for the \textit{Mapping Instruction to UI Action} task, it is worth noting that our method still achieved competitive accuracy with only two examples instead of extensive modeling on dedicated datasets like the benchmark. One possible reason for the underperformance is that LLMs are trained to generate text instead of selecting (predicting) an element id, and finetuning LLMs can potentially significantly boost the accuracy of this task. 
 As the first work in this direction, our work offers a set of new techniques and findings for using LLMs to enable conversational interaction. A natural next step is further examining these techniques by integrating them into an actual conversational agent. }

\subsection{Extending to Multi-Screen, Multi-Turn Conversational Interaction}
\label{multiturn}
\bw{
 We focus on unit conversation tasks involving single-turn interactions in this work. Future research can build upon our techniques to enable multi-screen \cite{motif} and multi-turn \cite{mug} conversational interactions. Multi-screen interactions can allow the agent to carry out complex tasks across various screens. For example, to help a user book a hotel room, an agent could start by asking which hotel the user wants to book and then search for that hotel (Figure \ref{dspace}-1: \textit{Screen Question-Generation}). Once the search results are obtained, the agent could announce a summary of the search result page (Figure \ref{dspace}-2: \textit{Screen Summary}). After that, the user could then ask for specific information from the search results, such as the room rates (Figure \ref{dspace}-3: \textit{Screen Question-Answering}), and finally, the user could command the agent to book the desired room (Figure \ref{dspace}-4: \textit{Mapping Instruction to UI Actions}). Multi-turn interactions can enable the agent to ask for clarifications on ambiguous cases \cite{li2020multi} or recover from errors. For example, given the screen in Figure \ref{tasks} with two emails sent from Google Play, if the user says, \textit{"Open the email sent by Google Play"}, our current approach would likely select one of them directly without clarifying with the users. A sophisticated agent should identify ambiguity and ask the user for clarification, such as, \textit{"Do you mean the one sent on Dec. 4th or Nov. 12th?"} To achieve multi-screen and multi-turn interactions with the proposed approach, future work will need to investigate methods to store and represent past screens and interactions as contexts \cite{openai_2022}}.

\subsection{Shots, Input Lengths, and Model Performance}
\label{shot-length}
Our findings indicate that the first prompt exemplar usually significantly boosts LLM performance in mobile UI tasks. In contrast, the inclusion of a second example only results in a marginal improvement. Previous research by Reynolds and McDonell \cite{reynolds2021prompt} suggests that the effectiveness of few-shot prompting lies within guiding LLMs to locate specific task locations in the model's existing space of learned tasks. Therefore, the first shot may be the most helpful, and additional examples may provide only marginal benefits in narrowing the model's focus. Relevantly, language models often have input length constraints, which limit the number of exemplars that can be included in the prompt. The length of a screen HTML has a considerable variance, depending on how much information was conveyed through the view hierarchy and the inherent complexity of the screen. Future work could explore approaches to prevent exceeding the input length limit. One potential approach is to dynamically select prompt screens with different lengths. When the test screen HTML is lengthy, shorter prompt screens could be selected. However, it is unclear whether imbalanced lengths between prompt and test screens would lead to inferior performances. Future research could investigate the trade-off between the number of shots, the length of each shot, and their impact on the model's performance in different UI tasks.

As modeling techniques advance, the input length restrictions of language models would likely be lifted \cite{lengthtransformer}, allowing for more information to be included in prompts. Another direction for future work is to develop methods for condensing screen HTML into a more concise syntax while still achieving comparable performance. In addition, to enable real-time applications, future research may investigate strategies such as model distillation \cite{hinton2015distill} and model compression \cite{song2015} for more efficient inference of LLMs.

\subsection{Screen Representation}
\label{screen_rep}
Mobile UI screens contain multiple modalities, including pixels, texts, and even audio when media content is present. A limitation of our investigation is that we only use the view hierarchy information, which is converted to an HTML representation, and leave other modalities unused. This limitation is imposed by the type of input expected by LLMs. While our studies showed LLMs could perform decently on various UI tasks, they could fail in cases requiring information not present in the view hierarchy but available as pixels. For instance, many icons or images on UI screens have missing captions or alt texts (text description of a visual element), and LLMs may not be able to perform tasks based on these elements. Moreover, visual information is particularly crucial for some apps, e.g., photo editing tools, and our approach may fall short of enabling conversational interaction based on these apps. Many models have started using multiple modalities, including visual and text information of UIs~\cite{zhang2021screen, 10.1145/3472749.3474765, li2020widget, motif}. Future work could exploit these prior models to generate missing captions or alt-text of elements, which can lead to more comprehensive screen information in the HTML input to LLMs. Our approach can also be extended by leveraging large-scale vision language models such as Flamingo \cite{Alayrac2022flamingo} to encode a screen's visual and structural information for few-shot learning.

\subsection{Steerability and Reliability of LLM Predictions}
Our experiments uncovered several promising capabilities of LLMs for mobile UI tasks. For example, the model showcased the behavior of question combination, i.e., merging relevant input fields into a single question, in the \textit{Screen Question-Generation} task. Furthermore, our study shows that the model can utilize its embedded prior knowledge to provide additional information in the \textit{Screen Summarization} task. While these capabilities are intriguing and potentially useful, there is currently a lack of direct controls for steering LLMs in terms of when and how these behaviors should occur. Additionally, LLMs, as other natural language generation models, could sometimes produce unintended text, known as hallucinations \cite{10.1145/3571730}. This means that the model may produce incorrect or irrelevant information. Our current method does not explicitly address this issue. As the research surrounding LLMs develops, it is crucial to find ways to improve the steerability and reliability of their predictions and behaviors, especially when incorporating LLMs into user-facing technology.

\subsection{Generalizing Beyond Mobile UIs}
This paper focuses on mobile UIs, but the proposed approach with LLMs can also be applied to other UI types, such as web UIs \cite{adept} (which are already in HTML syntax) and popular UI systems like iOS and macOS that possess view hierarchy data or equivalence. However, for more sophisticated and feature-rich UIs like video editors, a direct adaptation of our approach may be challenging due to the difficulty in representing key features in text format and ensuring the representation complies with model length constraints, as highlighted in the previous discussions. Nonetheless, our research provides a solid foundation for further studies to build upon and develop innovative methods that enable language interaction with graphical user interfaces.

\section{Conclusion}
We investigated the feasibility of prompting LLMs to enable various conversational interactions on mobile UIs. By categorizing the conversation scenarios between users and agents during mobile tasks, we identified four crucial UI tasks to study. We proposed a suite of prompting techniques for adapting LLMs to mobile UIs. We conducted extensive experiments with the four selected tasks to evaluate the effectiveness of our approach. The results showed that compared to traditional machine learning pipelines that consist of expensive data collection and model training, one could rapidly realize novel language-based interactions using LLMs while achieving competitive performance.
\begin{acks}
We thank the anonymous reviewers for their valuable feedback during the R\&R process that enhanced the paper's quality. We appreciate the discussions and feedback from our team members Chin-Yi Cheng and Tao Li. We also acknowledge the early-stage feedback from Michael Terry and Minsuk Chang. Special thanks to the Google Data Compute team for their invaluable assistance in data collection.
\end{acks}

\bibliographystyle{ACM-Reference-Format}
\bibliography{references.bib}


\begin{thebibliography}{59}


\ifx \showCODEN    \undefined \def \showCODEN     #1{\unskip}     \fi
\ifx \showDOI      \undefined \def \showDOI       #1{#1}\fi
\ifx \showISBNx    \undefined \def \showISBNx     #1{\unskip}     \fi
\ifx \showISBNxiii \undefined \def \showISBNxiii  #1{\unskip}     \fi
\ifx \showISSN     \undefined \def \showISSN      #1{\unskip}     \fi
\ifx \showLCCN     \undefined \def \showLCCN      #1{\unskip}     \fi
\ifx \shownote     \undefined \def \shownote      #1{#1}          \fi
\ifx \showarticletitle \undefined \def \showarticletitle #1{#1}   \fi
\ifx \showURL      \undefined \def \showURL       {\relax}        \fi
\providecommand\bibfield[2]{#2}
\providecommand\bibinfo[2]{#2}
\providecommand\natexlab[1]{#1}
\providecommand\showeprint[2][]{arXiv:#2}

\bibitem[Adept(2022)]%
        {adept}
\bibfield{author}{\bibinfo{person}{Adept}.} \bibinfo{year}{2022}\natexlab{}.
\newblock \bibinfo{booktitle}{\emph{ACT-1: Transformer for Actions}}.
\newblock
\urldef\tempurl%
\url{https://www.adept.ai/act}
\showURL{%
\tempurl}


\bibitem[Ahn et~al\mbox{.}(2022)]%
        {2022saycan}
\bibfield{author}{\bibinfo{person}{Michael Ahn}, \bibinfo{person}{Anthony
  Brohan}, \bibinfo{person}{Noah Brown}, \bibinfo{person}{Yevgen Chebotar},
  \bibinfo{person}{Omar Cortes}, \bibinfo{person}{Byron David},
  \bibinfo{person}{Chelsea Finn}, \bibinfo{person}{Chuyuan Fu},
  \bibinfo{person}{Keerthana Gopalakrishnan}, \bibinfo{person}{Karol Hausman},
  \bibinfo{person}{Alex Herzog}, \bibinfo{person}{Daniel Ho},
  \bibinfo{person}{Jasmine Hsu}, \bibinfo{person}{Julian Ibarz},
  \bibinfo{person}{Brian Ichter}, \bibinfo{person}{Alex Irpan},
  \bibinfo{person}{Eric Jang}, \bibinfo{person}{Rosario~Jauregui Ruano},
  \bibinfo{person}{Kyle Jeffrey}, \bibinfo{person}{Sally Jesmonth},
  \bibinfo{person}{Nikhil~J Joshi}, \bibinfo{person}{Ryan Julian},
  \bibinfo{person}{Dmitry Kalashnikov}, \bibinfo{person}{Yuheng Kuang},
  \bibinfo{person}{Kuang-Huei Lee}, \bibinfo{person}{Sergey Levine},
  \bibinfo{person}{Yao Lu}, \bibinfo{person}{Linda Luu},
  \bibinfo{person}{Carolina Parada}, \bibinfo{person}{Peter Pastor},
  \bibinfo{person}{Jornell Quiambao}, \bibinfo{person}{Kanishka Rao},
  \bibinfo{person}{Jarek Rettinghouse}, \bibinfo{person}{Diego Reyes},
  \bibinfo{person}{Pierre Sermanet}, \bibinfo{person}{Nicolas Sievers},
  \bibinfo{person}{Clayton Tan}, \bibinfo{person}{Alexander Toshev},
  \bibinfo{person}{Vincent Vanhoucke}, \bibinfo{person}{Fei Xia},
  \bibinfo{person}{Ted Xiao}, \bibinfo{person}{Peng Xu},
  \bibinfo{person}{Sichun Xu}, \bibinfo{person}{Mengyuan Yan}, {and}
  \bibinfo{person}{Andy Zeng}.} \bibinfo{year}{2022}\natexlab{}.
\newblock \bibinfo{title}{Do As I Can, Not As I Say: Grounding Language in
  Robotic Affordances}.
\newblock
\newblock
\urldef\tempurl%
\url{https://doi.org/10.48550/ARXIV.2204.01691}
\showDOI{\tempurl}


\bibitem[Alayrac et~al\mbox{.}(2022a)]%
        {flamingo}
\bibfield{author}{\bibinfo{person}{Jean-Baptiste Alayrac},
  \bibinfo{person}{Jeff Donahue}, \bibinfo{person}{Pauline Luc},
  \bibinfo{person}{Antoine Miech}, \bibinfo{person}{Iain Barr},
  \bibinfo{person}{Yana Hasson}, \bibinfo{person}{Karel Lenc},
  \bibinfo{person}{Arthur Mensch}, \bibinfo{person}{Katie Millican},
  \bibinfo{person}{Malcolm Reynolds}, \bibinfo{person}{Roman Ring},
  \bibinfo{person}{Eliza Rutherford}, \bibinfo{person}{Serkan Cabi},
  \bibinfo{person}{Tengda Han}, \bibinfo{person}{Zhitao Gong},
  \bibinfo{person}{Sina Samangooei}, \bibinfo{person}{Marianne Monteiro},
  \bibinfo{person}{Jacob Menick}, \bibinfo{person}{Sebastian Borgeaud},
  \bibinfo{person}{Andrew Brock}, \bibinfo{person}{Aida Nematzadeh},
  \bibinfo{person}{Sahand Sharifzadeh}, \bibinfo{person}{Mikolaj Binkowski},
  \bibinfo{person}{Ricardo Barreira}, \bibinfo{person}{Oriol Vinyals},
  \bibinfo{person}{Andrew Zisserman}, {and} \bibinfo{person}{Karen Simonyan}.}
  \bibinfo{year}{2022}\natexlab{a}.
\newblock \bibinfo{title}{Flamingo: a Visual Language Model for Few-Shot
  Learning}.
\newblock
\newblock
\urldef\tempurl%
\url{https://doi.org/10.48550/ARXIV.2204.14198}
\showDOI{\tempurl}


\bibitem[Alayrac et~al\mbox{.}(2022b)]%
        {Alayrac2022flamingo}
\bibfield{author}{\bibinfo{person}{Jean-Baptiste Alayrac},
  \bibinfo{person}{Jeff Donahue}, \bibinfo{person}{Pauline Luc},
  \bibinfo{person}{Antoine Miech}, \bibinfo{person}{Iain Barr},
  \bibinfo{person}{Yana Hasson}, \bibinfo{person}{Karel Lenc},
  \bibinfo{person}{Arthur Mensch}, \bibinfo{person}{Katie Millican},
  \bibinfo{person}{Malcolm Reynolds}, \bibinfo{person}{Roman Ring},
  \bibinfo{person}{Eliza Rutherford}, \bibinfo{person}{Serkan Cabi},
  \bibinfo{person}{Tengda Han}, \bibinfo{person}{Zhitao Gong},
  \bibinfo{person}{Sina Samangooei}, \bibinfo{person}{Marianne Monteiro},
  \bibinfo{person}{Jacob Menick}, \bibinfo{person}{Sebastian Borgeaud},
  \bibinfo{person}{Andrew Brock}, \bibinfo{person}{Aida Nematzadeh},
  \bibinfo{person}{Sahand Sharifzadeh}, \bibinfo{person}{Mikolaj Binkowski},
  \bibinfo{person}{Ricardo Barreira}, \bibinfo{person}{Oriol Vinyals},
  \bibinfo{person}{Andrew Zisserman}, {and} \bibinfo{person}{Karen Simonyan}.}
  \bibinfo{year}{2022}\natexlab{b}.
\newblock \bibinfo{title}{Flamingo: a Visual Language Model for Few-Shot
  Learning}.
\newblock
\newblock
\urldef\tempurl%
\url{https://doi.org/10.48550/ARXIV.2204.14198}
\showDOI{\tempurl}


\bibitem[Antol et~al\mbox{.}(2015)]%
        {antol2015vqa}
\bibfield{author}{\bibinfo{person}{Stanislaw Antol}, \bibinfo{person}{Aishwarya
  Agrawal}, \bibinfo{person}{Jiasen Lu}, \bibinfo{person}{Margaret Mitchell},
  \bibinfo{person}{Dhruv Batra}, \bibinfo{person}{C~Lawrence Zitnick}, {and}
  \bibinfo{person}{Devi Parikh}.} \bibinfo{year}{2015}\natexlab{}.
\newblock \showarticletitle{Vqa: Visual question answering}. In
  \bibinfo{booktitle}{\emph{Proceedings of the IEEE international conference on
  computer vision}}. \bibinfo{pages}{2425--2433}.
\newblock


\bibitem[Bolt(1980)]%
        {bolt1980put}
\bibfield{author}{\bibinfo{person}{Richard~A Bolt}.}
  \bibinfo{year}{1980}\natexlab{}.
\newblock \showarticletitle{“Put-that-there” Voice and gesture at the
  graphics interface}. In \bibinfo{booktitle}{\emph{Proceedings of the 7th
  annual conference on Computer graphics and interactive techniques}}.
  \bibinfo{pages}{262--270}.
\newblock


\bibitem[Brown et~al\mbox{.}(2020)]%
        {brown2020language}
\bibfield{author}{\bibinfo{person}{Tom~B. Brown}, \bibinfo{person}{Benjamin
  Mann}, \bibinfo{person}{Nick Ryder}, \bibinfo{person}{Melanie Subbiah},
  \bibinfo{person}{Jared Kaplan}, \bibinfo{person}{Prafulla Dhariwal},
  \bibinfo{person}{Arvind Neelakantan}, \bibinfo{person}{Pranav Shyam},
  \bibinfo{person}{Girish Sastry}, \bibinfo{person}{Amanda Askell},
  \bibinfo{person}{Sandhini Agarwal}, \bibinfo{person}{Ariel Herbert-Voss},
  \bibinfo{person}{Gretchen Krueger}, \bibinfo{person}{Tom Henighan},
  \bibinfo{person}{Rewon Child}, \bibinfo{person}{Aditya Ramesh},
  \bibinfo{person}{Daniel~M. Ziegler}, \bibinfo{person}{Jeffrey Wu},
  \bibinfo{person}{Clemens Winter}, \bibinfo{person}{Christopher Hesse},
  \bibinfo{person}{Mark Chen}, \bibinfo{person}{Eric Sigler},
  \bibinfo{person}{Mateusz Litwin}, \bibinfo{person}{Scott Gray},
  \bibinfo{person}{Benjamin Chess}, \bibinfo{person}{Jack Clark},
  \bibinfo{person}{Christopher Berner}, \bibinfo{person}{Sam McCandlish},
  \bibinfo{person}{Alec Radford}, \bibinfo{person}{Ilya Sutskever}, {and}
  \bibinfo{person}{Dario Amodei}.} \bibinfo{year}{2020}\natexlab{}.
\newblock \bibinfo{title}{Language Models are Few-Shot Learners}.
\newblock
\newblock
\showeprint[arxiv]{2005.14165}~[cs.CL]


\bibitem[Burns et~al\mbox{.}(2022)]%
        {motif}
\bibfield{author}{\bibinfo{person}{Andrea Burns}, \bibinfo{person}{Deniz
  Arsan}, \bibinfo{person}{Sanjna Agrawal}, \bibinfo{person}{Ranjitha Kumar},
  \bibinfo{person}{Kate Saenko}, {and} \bibinfo{person}{Bryan~A. Plummer}.}
  \bibinfo{year}{2022}\natexlab{}.
\newblock \bibinfo{title}{A Dataset for Interactive Vision-Language Navigation
  with Unknown Command Feasibility}.
\newblock
\newblock
\urldef\tempurl%
\url{https://doi.org/10.48550/ARXIV.2202.02312}
\showDOI{\tempurl}


\bibitem[Chowdhery et~al\mbox{.}(2022)]%
        {chowdhery2022palm}
\bibfield{author}{\bibinfo{person}{Aakanksha Chowdhery},
  \bibinfo{person}{Sharan Narang}, \bibinfo{person}{Jacob Devlin},
  \bibinfo{person}{Maarten Bosma}, \bibinfo{person}{Gaurav Mishra},
  \bibinfo{person}{Adam Roberts}, \bibinfo{person}{Paul Barham},
  \bibinfo{person}{Hyung~Won Chung}, \bibinfo{person}{Charles Sutton},
  \bibinfo{person}{Sebastian Gehrmann}, {et~al\mbox{.}}}
  \bibinfo{year}{2022}\natexlab{}.
\newblock \showarticletitle{Palm: Scaling language modeling with pathways}.
\newblock \bibinfo{journal}{\emph{arXiv preprint arXiv:2204.02311}}
  (\bibinfo{year}{2022}).
\newblock


\bibitem[Chung et~al\mbox{.}(2022)]%
        {chung2022talebrush}
\bibfield{author}{\bibinfo{person}{John Joon~Young Chung},
  \bibinfo{person}{Wooseok Kim}, \bibinfo{person}{Kang~Min Yoo},
  \bibinfo{person}{Hwaran Lee}, \bibinfo{person}{Eytan Adar}, {and}
  \bibinfo{person}{Minsuk Chang}.} \bibinfo{year}{2022}\natexlab{}.
\newblock \showarticletitle{TaleBrush: Visual Sketching of Story Generation
  with Pretrained Language Models}. In \bibinfo{booktitle}{\emph{Extended
  Abstracts of the 2022 CHI Conference on Human Factors in Computing Systems}}
  (New Orleans, LA, USA) \emph{(\bibinfo{series}{CHI EA '22})}.
  \bibinfo{publisher}{Association for Computing Machinery},
  \bibinfo{address}{New York, NY, USA}, Article \bibinfo{articleno}{172},
  \bibinfo{numpages}{4}~pages.
\newblock
\showISBNx{9781450391566}
\urldef\tempurl%
\url{https://doi.org/10.1145/3491101.3519873}
\showDOI{\tempurl}


\bibitem[Dang et~al\mbox{.}(2022)]%
        {dang2022prompt}
\bibfield{author}{\bibinfo{person}{Hai Dang}, \bibinfo{person}{Lukas Mecke},
  \bibinfo{person}{Florian Lehmann}, \bibinfo{person}{Sven Goller}, {and}
  \bibinfo{person}{Daniel Buschek}.} \bibinfo{year}{2022}\natexlab{}.
\newblock \bibinfo{title}{How to Prompt? Opportunities and Challenges of Zero-
  and Few-Shot Learning for Human-AI Interaction in Creative Applications of
  Generative Models}.
\newblock
\newblock
\urldef\tempurl%
\url{https://doi.org/10.48550/ARXIV.2209.01390}
\showDOI{\tempurl}


\bibitem[Deka et~al\mbox{.}(2017)]%
        {Deka:2017:Rico}
\bibfield{author}{\bibinfo{person}{Biplab Deka}, \bibinfo{person}{Zifeng
  Huang}, \bibinfo{person}{Chad Franzen}, \bibinfo{person}{Joshua Hibschman},
  \bibinfo{person}{Daniel Afergan}, \bibinfo{person}{Yang Li},
  \bibinfo{person}{Jeffrey Nichols}, {and} \bibinfo{person}{Ranjitha Kumar}.}
  \bibinfo{year}{2017}\natexlab{}.
\newblock \showarticletitle{Rico: A Mobile App Dataset for Building Data-Driven
  Design Applications}. In \bibinfo{booktitle}{\emph{Proceedings of the 30th
  Annual Symposium on User Interface Software and Technology}}
  \emph{(\bibinfo{series}{UIST '17})}.
\newblock


\bibitem[F{\o}lstad and Brandtz{\ae}g(2017)]%
        {folstad2017chatbots}
\bibfield{author}{\bibinfo{person}{Asbj{\o}rn F{\o}lstad} {and}
  \bibinfo{person}{Petter~Bae Brandtz{\ae}g}.} \bibinfo{year}{2017}\natexlab{}.
\newblock \showarticletitle{Chatbots and the new world of HCI}.
\newblock \bibinfo{journal}{\emph{interactions}} \bibinfo{volume}{24},
  \bibinfo{number}{4} (\bibinfo{year}{2017}), \bibinfo{pages}{38--42}.
\newblock


\bibitem[Goyal et~al\mbox{.}(2022)]%
        {gptsummarization}
\bibfield{author}{\bibinfo{person}{Tanya Goyal}, \bibinfo{person}{Junyi~Jessy
  Li}, {and} \bibinfo{person}{Greg Durrett}.} \bibinfo{year}{2022}\natexlab{}.
\newblock \bibinfo{title}{News Summarization and Evaluation in the Era of
  GPT-3}.
\newblock
\newblock
\urldef\tempurl%
\url{https://doi.org/10.48550/ARXIV.2209.12356}
\showDOI{\tempurl}


\bibitem[Han et~al\mbox{.}(2015)]%
        {song2015}
\bibfield{author}{\bibinfo{person}{Song Han}, \bibinfo{person}{Huizi Mao},
  {and} \bibinfo{person}{William~J. Dally}.} \bibinfo{year}{2015}\natexlab{}.
\newblock \bibinfo{title}{Deep Compression: Compressing Deep Neural Networks
  with Pruning, Trained Quantization and Huffman Coding}.
\newblock
\newblock
\urldef\tempurl%
\url{https://doi.org/10.48550/ARXIV.1510.00149}
\showDOI{\tempurl}


\bibitem[Hinton et~al\mbox{.}(2015)]%
        {hinton2015distill}
\bibfield{author}{\bibinfo{person}{Geoffrey Hinton}, \bibinfo{person}{Oriol
  Vinyals}, {and} \bibinfo{person}{Jeff Dean}.}
  \bibinfo{year}{2015}\natexlab{}.
\newblock \bibinfo{title}{Distilling the Knowledge in a Neural Network}.
\newblock
\newblock
\urldef\tempurl%
\url{https://doi.org/10.48550/ARXIV.1503.02531}
\showDOI{\tempurl}


\bibitem[Horvitz(1999)]%
        {horvitz1999principles}
\bibfield{author}{\bibinfo{person}{Eric Horvitz}.}
  \bibinfo{year}{1999}\natexlab{}.
\newblock \showarticletitle{Principles of mixed-initiative user interfaces}. In
  \bibinfo{booktitle}{\emph{Proceedings of the SIGCHI conference on Human
  Factors in Computing Systems}}. \bibinfo{pages}{159--166}.
\newblock


\bibitem[Hsiao et~al\mbox{.}(2022)]%
        {screenqa}
\bibfield{author}{\bibinfo{person}{Yu-Chung Hsiao}, \bibinfo{person}{Fedir
  Zubach}, \bibinfo{person}{Maria Wang}, {and} \bibinfo{person}{Chen Jindong}.}
  \bibinfo{year}{2022}\natexlab{}.
\newblock \bibinfo{title}{ScreenQA: Large-Scale Question-Answer Pairs over
  Mobile App Screenshots}.
\newblock
\newblock
\urldef\tempurl%
\url{https://doi.org/10.48550/ARXIV.2209.08199}
\showDOI{\tempurl}


\bibitem[Ji et~al\mbox{.}(2022)]%
        {10.1145/3571730}
\bibfield{author}{\bibinfo{person}{Ziwei Ji}, \bibinfo{person}{Nayeon Lee},
  \bibinfo{person}{Rita Frieske}, \bibinfo{person}{Tiezheng Yu},
  \bibinfo{person}{Dan Su}, \bibinfo{person}{Yan Xu}, \bibinfo{person}{Etsuko
  Ishii}, \bibinfo{person}{Yejin Bang}, \bibinfo{person}{Andrea Madotto}, {and}
  \bibinfo{person}{Pascale Fung}.} \bibinfo{year}{2022}\natexlab{}.
\newblock \showarticletitle{Survey of Hallucination in Natural Language
  Generation}.
\newblock \bibinfo{journal}{\emph{ACM Comput. Surv.}} (\bibinfo{date}{nov}
  \bibinfo{year}{2022}).
\newblock
\showISSN{0360-0300}
\urldef\tempurl%
\url{https://doi.org/10.1145/3571730}
\showDOI{\tempurl}
\newblock
\shownote{Just Accepted}.


\bibitem[Jiang et~al\mbox{.}(2022a)]%
        {jian2022case}
\bibfield{author}{\bibinfo{person}{Ellen Jiang}, \bibinfo{person}{Kristen
  Olson}, \bibinfo{person}{Edwin Toh}, \bibinfo{person}{Alejandra Molina},
  \bibinfo{person}{Aaron Donsbach}, \bibinfo{person}{Michael Terry}, {and}
  \bibinfo{person}{Carrie~J Cai}.} \bibinfo{year}{2022}\natexlab{a}.
\newblock \showarticletitle{PromptMaker: Prompt-Based Prototyping with Large
  Language Models}. In \bibinfo{booktitle}{\emph{Extended Abstracts of the 2022
  CHI Conference on Human Factors in Computing Systems}} (New Orleans, LA, USA)
  \emph{(\bibinfo{series}{CHI EA '22})}. \bibinfo{publisher}{Association for
  Computing Machinery}, \bibinfo{address}{New York, NY, USA}, Article
  \bibinfo{articleno}{35}, \bibinfo{numpages}{8}~pages.
\newblock
\showISBNx{9781450391566}
\urldef\tempurl%
\url{https://doi.org/10.1145/3491101.3503564}
\showDOI{\tempurl}


\bibitem[Jiang et~al\mbox{.}(2022b)]%
        {jiang2022nlp}
\bibfield{author}{\bibinfo{person}{Ellen Jiang}, \bibinfo{person}{Edwin Toh},
  \bibinfo{person}{Alejandra Molina}, \bibinfo{person}{Kristen Olson},
  \bibinfo{person}{Claire Kayacik}, \bibinfo{person}{Aaron Donsbach},
  \bibinfo{person}{Carrie~J Cai}, {and} \bibinfo{person}{Michael Terry}.}
  \bibinfo{year}{2022}\natexlab{b}.
\newblock \showarticletitle{Discovering the Syntax and Strategies of Natural
  Language Programming with Generative Language Models}. In
  \bibinfo{booktitle}{\emph{Proceedings of the 2022 CHI Conference on Human
  Factors in Computing Systems}} (New Orleans, LA, USA)
  \emph{(\bibinfo{series}{CHI '22})}. \bibinfo{publisher}{Association for
  Computing Machinery}, \bibinfo{address}{New York, NY, USA}, Article
  \bibinfo{articleno}{386}, \bibinfo{numpages}{19}~pages.
\newblock
\showISBNx{9781450391573}
\urldef\tempurl%
\url{https://doi.org/10.1145/3491102.3501870}
\showDOI{\tempurl}


\bibitem[Karat et~al\mbox{.}(2002)]%
        {karat2002conversational}
\bibfield{author}{\bibinfo{person}{Clare-Marie Karat}, \bibinfo{person}{John
  Vergo}, {and} \bibinfo{person}{David Nahamoo}.}
  \bibinfo{year}{2002}\natexlab{}.
\newblock \showarticletitle{Conversational interface technologies}.
\newblock In \bibinfo{booktitle}{\emph{The human-computer interaction handbook:
  fundamentals, evolving technologies and emerging applications}}.
  \bibinfo{pages}{169--186}.
\newblock


\bibitem[Kim et~al\mbox{.}(2022)]%
        {kim2022stylette}
\bibfield{author}{\bibinfo{person}{Tae~Soo Kim}, \bibinfo{person}{DaEun Choi},
  \bibinfo{person}{Yoonseo Choi}, {and} \bibinfo{person}{Juho Kim}.}
  \bibinfo{year}{2022}\natexlab{}.
\newblock \showarticletitle{Stylette: Styling the Web with Natural Language}.
  In \bibinfo{booktitle}{\emph{Proceedings of the 2022 CHI Conference on Human
  Factors in Computing Systems}} (New Orleans, LA, USA)
  \emph{(\bibinfo{series}{CHI '22})}. \bibinfo{publisher}{Association for
  Computing Machinery}, \bibinfo{address}{New York, NY, USA}, Article
  \bibinfo{articleno}{5}, \bibinfo{numpages}{17}~pages.
\newblock
\showISBNx{9781450391573}
\urldef\tempurl%
\url{https://doi.org/10.1145/3491102.3501931}
\showDOI{\tempurl}


\bibitem[Kojima et~al\mbox{.}(2022)]%
        {kojima2022think}
\bibfield{author}{\bibinfo{person}{Takeshi Kojima},
  \bibinfo{person}{Shixiang~Shane Gu}, \bibinfo{person}{Machel Reid},
  \bibinfo{person}{Yutaka Matsuo}, {and} \bibinfo{person}{Yusuke Iwasawa}.}
  \bibinfo{year}{2022}\natexlab{}.
\newblock \bibinfo{title}{Large Language Models are Zero-Shot Reasoners}.
\newblock
\newblock
\urldef\tempurl%
\url{https://doi.org/10.48550/ARXIV.2205.11916}
\showDOI{\tempurl}


\bibitem[Lee et~al\mbox{.}(2022c)]%
        {lee2022coauthor}
\bibfield{author}{\bibinfo{person}{Mina Lee}, \bibinfo{person}{Percy Liang},
  {and} \bibinfo{person}{Qian Yang}.} \bibinfo{year}{2022}\natexlab{c}.
\newblock \showarticletitle{CoAuthor: Designing a Human-AI Collaborative
  Writing Dataset for Exploring Language Model Capabilities}. In
  \bibinfo{booktitle}{\emph{Proceedings of the 2022 CHI Conference on Human
  Factors in Computing Systems}} (New Orleans, LA, USA)
  \emph{(\bibinfo{series}{CHI '22})}. \bibinfo{publisher}{Association for
  Computing Machinery}, \bibinfo{address}{New York, NY, USA}, Article
  \bibinfo{articleno}{388}, \bibinfo{numpages}{19}~pages.
\newblock
\showISBNx{9781450391573}
\urldef\tempurl%
\url{https://doi.org/10.1145/3491102.3502030}
\showDOI{\tempurl}


\bibitem[Lee et~al\mbox{.}(2022a)]%
        {lee2022promptiverse}
\bibfield{author}{\bibinfo{person}{Yoonjoo Lee}, \bibinfo{person}{John
  Joon~Young Chung}, \bibinfo{person}{Tae~Soo Kim}, \bibinfo{person}{Jean~Y
  Song}, {and} \bibinfo{person}{Juho Kim}.} \bibinfo{year}{2022}\natexlab{a}.
\newblock \showarticletitle{Promptiverse: Scalable Generation of Scaffolding
  Prompts Through Human-AI Hybrid Knowledge Graph Annotation}. In
  \bibinfo{booktitle}{\emph{Proceedings of the 2022 CHI Conference on Human
  Factors in Computing Systems}} (New Orleans, LA, USA)
  \emph{(\bibinfo{series}{CHI '22})}. \bibinfo{publisher}{Association for
  Computing Machinery}, \bibinfo{address}{New York, NY, USA}, Article
  \bibinfo{articleno}{96}, \bibinfo{numpages}{18}~pages.
\newblock
\showISBNx{9781450391573}
\urldef\tempurl%
\url{https://doi.org/10.1145/3491102.3502087}
\showDOI{\tempurl}


\bibitem[Lee et~al\mbox{.}(2022b)]%
        {lee2022interactive}
\bibfield{author}{\bibinfo{person}{Yoonjoo Lee}, \bibinfo{person}{Tae~Soo Kim},
  \bibinfo{person}{Minsuk Chang}, {and} \bibinfo{person}{Juho Kim}.}
  \bibinfo{year}{2022}\natexlab{b}.
\newblock \showarticletitle{Interactive Children’s Story Rewriting Through
  Parent-Children Interaction}. In \bibinfo{booktitle}{\emph{Proceedings of the
  First Workshop on Intelligent and Interactive Writing Assistants (In2Writing
  2022)}}. \bibinfo{pages}{62--71}.
\newblock


\bibitem[Leiva et~al\mbox{.}(2022)]%
        {10.1145/3564702}
\bibfield{author}{\bibinfo{person}{Luis~A. Leiva}, \bibinfo{person}{Asutosh
  Hota}, {and} \bibinfo{person}{Antti Oulasvirta}.}
  \bibinfo{year}{2022}\natexlab{}.
\newblock \showarticletitle{Describing UI Screenshots in Natural Language}.
\newblock \bibinfo{journal}{\emph{ACM Trans. Intell. Syst. Technol.}}
  \bibinfo{volume}{14}, \bibinfo{number}{1}, Article \bibinfo{articleno}{19}
  (\bibinfo{date}{nov} \bibinfo{year}{2022}), \bibinfo{numpages}{28}~pages.
\newblock
\showISSN{2157-6904}
\urldef\tempurl%
\url{https://doi.org/10.1145/3564702}
\showDOI{\tempurl}


\bibitem[Li et~al\mbox{.}(2022a)]%
        {li2022denoising}
\bibfield{author}{\bibinfo{person}{Gang Li}, \bibinfo{person}{Gilles Baechler},
  \bibinfo{person}{Manuel Tragut}, {and} \bibinfo{person}{Yang Li}.}
  \bibinfo{year}{2022}\natexlab{a}.
\newblock \showarticletitle{Learning to Denoise Raw Mobile UI Layouts for
  Improving Datasets at Scale}. In \bibinfo{booktitle}{\emph{Proceedings of the
  2022 CHI Conference on Human Factors in Computing Systems}} (New Orleans, LA,
  USA) \emph{(\bibinfo{series}{CHI '22})}. \bibinfo{publisher}{Association for
  Computing Machinery}, \bibinfo{address}{New York, NY, USA}, Article
  \bibinfo{articleno}{67}, \bibinfo{numpages}{13}~pages.
\newblock
\showISBNx{9781450391573}
\urldef\tempurl%
\url{https://doi.org/10.1145/3491102.3502042}
\showDOI{\tempurl}


\bibitem[Li et~al\mbox{.}(2022b)]%
        {mug}
\bibfield{author}{\bibinfo{person}{Tao Li}, \bibinfo{person}{Gang Li},
  \bibinfo{person}{Jingjie Zheng}, \bibinfo{person}{Purple Wang}, {and}
  \bibinfo{person}{Yang Li}.} \bibinfo{year}{2022}\natexlab{b}.
\newblock \bibinfo{title}{MUG: Interactive Multimodal Grounding on User
  Interfaces}.
\newblock
\newblock
\urldef\tempurl%
\url{https://doi.org/10.48550/ARXIV.2209.15099}
\showDOI{\tempurl}


\bibitem[Li et~al\mbox{.}(2017)]%
        {li2017sugilite}
\bibfield{author}{\bibinfo{person}{Toby Jia-Jun Li}, \bibinfo{person}{Amos
  Azaria}, {and} \bibinfo{person}{Brad~A Myers}.}
  \bibinfo{year}{2017}\natexlab{}.
\newblock \showarticletitle{SUGILITE: creating multimodal smartphone automation
  by demonstration}. In \bibinfo{booktitle}{\emph{Proceedings of the 2017 CHI
  conference on human factors in computing systems}}.
  \bibinfo{pages}{6038--6049}.
\newblock


\bibitem[Li et~al\mbox{.}(2020a)]%
        {li2020multi}
\bibfield{author}{\bibinfo{person}{Toby Jia-Jun Li}, \bibinfo{person}{Jingya
  Chen}, \bibinfo{person}{Haijun Xia}, \bibinfo{person}{Tom~M Mitchell}, {and}
  \bibinfo{person}{Brad~A Myers}.} \bibinfo{year}{2020}\natexlab{a}.
\newblock \showarticletitle{Multi-modal repairs of conversational breakdowns in
  task-oriented dialogs}. In \bibinfo{booktitle}{\emph{Proceedings of the 33rd
  Annual ACM Symposium on User Interface Software and Technology}}.
  \bibinfo{pages}{1094--1107}.
\newblock


\bibitem[Li et~al\mbox{.}(2018)]%
        {8506506}
\bibfield{author}{\bibinfo{person}{Toby Jia-Jun Li}, \bibinfo{person}{Igor
  Labutov}, \bibinfo{person}{Xiaohan~Nancy Li}, \bibinfo{person}{Xiaoyi Zhang},
  \bibinfo{person}{Wenze Shi}, \bibinfo{person}{Wanling Ding},
  \bibinfo{person}{Tom~M. Mitchell}, {and} \bibinfo{person}{Brad~A. Myers}.}
  \bibinfo{year}{2018}\natexlab{}.
\newblock \showarticletitle{APPINITE: A Multi-Modal Interface for Specifying
  Data Descriptions in Programming by Demonstration Using Natural Language
  Instructions}. In \bibinfo{booktitle}{\emph{2018 IEEE Symposium on Visual
  Languages and Human-Centric Computing (VL/HCC)}}. \bibinfo{pages}{105--114}.
\newblock
\urldef\tempurl%
\url{https://doi.org/10.1109/VLHCC.2018.8506506}
\showDOI{\tempurl}


\bibitem[Li et~al\mbox{.}(2021c)]%
        {li2021screen2vec}
\bibfield{author}{\bibinfo{person}{Toby Jia-Jun Li}, \bibinfo{person}{Lindsay
  Popowski}, \bibinfo{person}{Tom Mitchell}, {and} \bibinfo{person}{Brad~A
  Myers}.} \bibinfo{year}{2021}\natexlab{c}.
\newblock \showarticletitle{Screen2vec: Semantic embedding of gui screens and
  gui components}. In \bibinfo{booktitle}{\emph{Proceedings of the 2021 CHI
  Conference on Human Factors in Computing Systems}}. \bibinfo{pages}{1--15}.
\newblock


\bibitem[Li et~al\mbox{.}(2019)]%
        {li2019pumice}
\bibfield{author}{\bibinfo{person}{Toby Jia-Jun Li}, \bibinfo{person}{Marissa
  Radensky}, \bibinfo{person}{Justin Jia}, \bibinfo{person}{Kirielle
  Singarajah}, \bibinfo{person}{Tom~M. Mitchell}, {and}
  \bibinfo{person}{Brad~A. Myers}.} \bibinfo{year}{2019}\natexlab{}.
\newblock \showarticletitle{PUMICE: A Multi-Modal Agent That Learns Concepts
  and Conditionals from Natural Language and Demonstrations}. In
  \bibinfo{booktitle}{\emph{Proceedings of the 32nd Annual ACM Symposium on
  User Interface Software and Technology}} (New Orleans, LA, USA)
  \emph{(\bibinfo{series}{UIST '19})}. \bibinfo{publisher}{Association for
  Computing Machinery}, \bibinfo{address}{New York, NY, USA},
  \bibinfo{pages}{577–589}.
\newblock
\showISBNx{9781450368162}
\urldef\tempurl%
\url{https://doi.org/10.1145/3332165.3347899}
\showDOI{\tempurl}


\bibitem[Li and Riva(2018)]%
        {li20218kite}
\bibfield{author}{\bibinfo{person}{Toby Jia-Jun Li} {and}
  \bibinfo{person}{Oriana Riva}.} \bibinfo{year}{2018}\natexlab{}.
\newblock \showarticletitle{Kite: Building Conversational Bots from Mobile
  Apps}. In \bibinfo{booktitle}{\emph{Proceedings of the 16th Annual
  International Conference on Mobile Systems, Applications, and Services}}
  (Munich, Germany) \emph{(\bibinfo{series}{MobiSys '18})}.
  \bibinfo{publisher}{Association for Computing Machinery},
  \bibinfo{address}{New York, NY, USA}, \bibinfo{pages}{96–109}.
\newblock
\showISBNx{9781450357203}
\urldef\tempurl%
\url{https://doi.org/10.1145/3210240.3210339}
\showDOI{\tempurl}


\bibitem[Li et~al\mbox{.}(2020b)]%
        {li-etal-2020-mapping}
\bibfield{author}{\bibinfo{person}{Yang Li}, \bibinfo{person}{Jiacong He},
  \bibinfo{person}{Xin Zhou}, \bibinfo{person}{Yuan Zhang}, {and}
  \bibinfo{person}{Jason Baldridge}.} \bibinfo{year}{2020}\natexlab{b}.
\newblock \showarticletitle{Mapping Natural Language Instructions to Mobile
  {UI} Action Sequences}. In \bibinfo{booktitle}{\emph{Proceedings of the 58th
  Annual Meeting of the Association for Computational Linguistics}}.
  \bibinfo{publisher}{Association for Computational Linguistics},
  \bibinfo{address}{Online}, \bibinfo{pages}{8198--8210}.
\newblock
\urldef\tempurl%
\url{https://doi.org/10.18653/v1/2020.acl-main.729}
\showDOI{\tempurl}


\bibitem[Li et~al\mbox{.}(2020c)]%
        {li2020widget}
\bibfield{author}{\bibinfo{person}{Yang Li}, \bibinfo{person}{Gang Li},
  \bibinfo{person}{Luheng He}, \bibinfo{person}{Jingjie Zheng},
  \bibinfo{person}{Hong Li}, {and} \bibinfo{person}{Zhiwei~(Wei) Guan}.}
  \bibinfo{year}{2020}\natexlab{c}.
\newblock \showarticletitle{Widget Captioning: Generating Natural Language
  Description for Mobile User Interface Elements}.
\newblock
\urldef\tempurl%
\url{https://www.aclweb.org/anthology/2020.emnlp-main.443.pdf}
\showURL{%
\tempurl}


\bibitem[Li et~al\mbox{.}(2021a)]%
        {li2021vut}
\bibfield{author}{\bibinfo{person}{Yang Li}, \bibinfo{person}{Gang Li},
  \bibinfo{person}{Xin Zhou}, \bibinfo{person}{Mostafa Dehghani}, {and}
  \bibinfo{person}{Alexey Gritsenko}.} \bibinfo{year}{2021}\natexlab{a}.
\newblock \showarticletitle{VUT: Versatile UI Transformer for Multi-Modal
  Multi-Task User Interface Modeling}.
\newblock \bibinfo{journal}{\emph{arXiv preprint arXiv:2112.05692}}
  (\bibinfo{year}{2021}).
\newblock


\bibitem[Li et~al\mbox{.}(2021b)]%
        {vut2021}
\bibfield{author}{\bibinfo{person}{Yang Li}, \bibinfo{person}{Gang Li},
  \bibinfo{person}{Xin Zhou}, \bibinfo{person}{Mostafa Dehghani}, {and}
  \bibinfo{person}{Alexey Gritsenko}.} \bibinfo{year}{2021}\natexlab{b}.
\newblock \bibinfo{title}{VUT: Versatile UI Transformer for Multi-Modal
  Multi-Task User Interface Modeling}.
\newblock
\newblock
\urldef\tempurl%
\url{https://doi.org/10.48550/ARXIV.2112.05692}
\showDOI{\tempurl}


\bibitem[Liu et~al\mbox{.}(2018)]%
        {liu2018reinforcement}
\bibfield{author}{\bibinfo{person}{Evan~Zheran Liu}, \bibinfo{person}{Kelvin
  Guu}, \bibinfo{person}{Panupong Pasupat}, \bibinfo{person}{Tianlin Shi},
  {and} \bibinfo{person}{Percy Liang}.} \bibinfo{year}{2018}\natexlab{}.
\newblock \bibinfo{title}{Reinforcement Learning on Web Interfaces Using
  Workflow-Guided Exploration}.
\newblock
\newblock
\showeprint[arxiv]{1802.08802}~[cs.AI]


\bibitem[Liu et~al\mbox{.}(2022)]%
        {liu2022will}
\bibfield{author}{\bibinfo{person}{Yihe Liu}, \bibinfo{person}{Anushk Mittal},
  \bibinfo{person}{Diyi Yang}, {and} \bibinfo{person}{Amy Bruckman}.}
  \bibinfo{year}{2022}\natexlab{}.
\newblock \showarticletitle{Will AI Console Me When I Lose My Pet?
  Understanding Perceptions of AI-Mediated Email Writing}. In
  \bibinfo{booktitle}{\emph{Proceedings of the 2022 CHI Conference on Human
  Factors in Computing Systems}} (New Orleans, LA, USA)
  \emph{(\bibinfo{series}{CHI '22})}. \bibinfo{publisher}{Association for
  Computing Machinery}, \bibinfo{address}{New York, NY, USA}, Article
  \bibinfo{articleno}{474}, \bibinfo{numpages}{13}~pages.
\newblock
\showISBNx{9781450391573}
\urldef\tempurl%
\url{https://doi.org/10.1145/3491102.3517731}
\showDOI{\tempurl}


\bibitem[OpenAI(2022)]%
        {openai_2022}
\bibfield{author}{\bibinfo{person}{OpenAI}.} \bibinfo{year}{2022}\natexlab{}.
\newblock \bibinfo{title}{CHATGPT: Optimizing language models for dialogue}.
\newblock
\newblock
\urldef\tempurl%
\url{https://openai.com/blog/chatgpt/}
\showURL{%
\tempurl}


\bibitem[Pasupat et~al\mbox{.}(2018)]%
        {pasupat-etal-2018-mapping}
\bibfield{author}{\bibinfo{person}{Panupong Pasupat},
  \bibinfo{person}{Tian-Shun Jiang}, \bibinfo{person}{Evan Liu},
  \bibinfo{person}{Kelvin Guu}, {and} \bibinfo{person}{Percy Liang}.}
  \bibinfo{year}{2018}\natexlab{}.
\newblock \showarticletitle{Mapping natural language commands to web elements}.
  In \bibinfo{booktitle}{\emph{Proceedings of the 2018 Conference on Empirical
  Methods in Natural Language Processing}}. \bibinfo{publisher}{Association for
  Computational Linguistics}, \bibinfo{address}{Brussels, Belgium},
  \bibinfo{pages}{4970--4976}.
\newblock
\urldef\tempurl%
\url{https://doi.org/10.18653/v1/D18-1540}
\showDOI{\tempurl}


\bibitem[Rajpurkar et~al\mbox{.}(2016)]%
        {rajpurkar2016squad}
\bibfield{author}{\bibinfo{person}{Pranav Rajpurkar}, \bibinfo{person}{Jian
  Zhang}, \bibinfo{person}{Konstantin Lopyrev}, {and} \bibinfo{person}{Percy
  Liang}.} \bibinfo{year}{2016}\natexlab{}.
\newblock \showarticletitle{Squad: 100,000+ questions for machine comprehension
  of text}.
\newblock \bibinfo{journal}{\emph{arXiv preprint arXiv:1606.05250}}
  (\bibinfo{year}{2016}).
\newblock


\bibitem[Reynolds and McDonell(2021)]%
        {reynolds2021prompt}
\bibfield{author}{\bibinfo{person}{Laria Reynolds} {and} \bibinfo{person}{Kyle
  McDonell}.} \bibinfo{year}{2021}\natexlab{}.
\newblock \bibinfo{title}{Prompt Programming for Large Language Models: Beyond
  the Few-Shot Paradigm}.
\newblock
\newblock
\urldef\tempurl%
\url{https://doi.org/10.48550/ARXIV.2102.07350}
\showDOI{\tempurl}


\bibitem[Sanh et~al\mbox{.}(2019)]%
        {Sanh2019distilbert}
\bibfield{author}{\bibinfo{person}{Victor Sanh}, \bibinfo{person}{Lysandre
  Debut}, \bibinfo{person}{Julien Chaumond}, {and} \bibinfo{person}{Thomas
  Wolf}.} \bibinfo{year}{2019}\natexlab{}.
\newblock \bibinfo{title}{DistilBERT, a distilled version of BERT: smaller,
  faster, cheaper and lighter}.
\newblock
\newblock
\urldef\tempurl%
\url{https://doi.org/10.48550/ARXIV.1910.01108}
\showDOI{\tempurl}


\bibitem[Sarsenbayeva et~al\mbox{.}(2017)]%
        {sarsenbayeva2017challenges}
\bibfield{author}{\bibinfo{person}{Zhanna Sarsenbayeva}, \bibinfo{person}{Niels
  van Berkel}, \bibinfo{person}{Chu Luo}, \bibinfo{person}{Vassilis Kostakos},
  {and} \bibinfo{person}{Jorge Goncalves}.} \bibinfo{year}{2017}\natexlab{}.
\newblock \showarticletitle{Challenges of situational impairments during
  interaction with mobile devices}. In \bibinfo{booktitle}{\emph{Proceedings of
  the 29th Australian Conference on Computer-Human Interaction}}.
  \bibinfo{pages}{477--481}.
\newblock


\bibitem[Todi et~al\mbox{.}(2021)]%
        {todi21conversations}
\bibfield{author}{\bibinfo{person}{Kashyap Todi}, \bibinfo{person}{Luis~A.
  Leiva}, \bibinfo{person}{Daniel Buschek}, \bibinfo{person}{Pin Tian}, {and}
  \bibinfo{person}{Antti Oulasvirta}.} \bibinfo{year}{2021}\natexlab{}.
\newblock \showarticletitle{{Conversations with GUIs}}. In
  \bibinfo{booktitle}{\emph{Proceedings of the ACM SIGCHI Conference on
  Designing Interactive Systems}} \emph{(\bibinfo{series}{DIS '21'})}.
  \bibinfo{publisher}{Association for Computing Machinery},
  \bibinfo{address}{New York, NY, USA}.
\newblock
\urldef\tempurl%
\url{https://doi.org/10.1145/3461778.3462124}
\showDOI{\tempurl}


\bibitem[Wang et~al\mbox{.}(2021)]%
        {10.1145/3472749.3474765}
\bibfield{author}{\bibinfo{person}{Bryan Wang}, \bibinfo{person}{Gang Li},
  \bibinfo{person}{Xin Zhou}, \bibinfo{person}{Zhourong Chen},
  \bibinfo{person}{Tovi Grossman}, {and} \bibinfo{person}{Yang Li}.}
  \bibinfo{year}{2021}\natexlab{}.
\newblock \showarticletitle{Screen2Words: Automatic Mobile UI Summarization
  with Multimodal Learning}. In \bibinfo{booktitle}{\emph{The 34th Annual ACM
  Symposium on User Interface Software and Technology}} (Virtual Event, USA)
  \emph{(\bibinfo{series}{UIST '21})}. \bibinfo{publisher}{Association for
  Computing Machinery}, \bibinfo{address}{New York, NY, USA},
  \bibinfo{pages}{498–510}.
\newblock
\showISBNx{9781450386357}
\urldef\tempurl%
\url{https://doi.org/10.1145/3472749.3474765}
\showDOI{\tempurl}


\bibitem[Wang et~al\mbox{.}(2022)]%
        {wang2022rae}
\bibfield{author}{\bibinfo{person}{Xuezhi Wang}, \bibinfo{person}{Jason Wei},
  \bibinfo{person}{Dale Schuurmans}, \bibinfo{person}{Quoc Le},
  \bibinfo{person}{Ed Chi}, {and} \bibinfo{person}{Denny Zhou}.}
  \bibinfo{year}{2022}\natexlab{}.
\newblock \bibinfo{title}{Rationale-Augmented Ensembles in Language Models}.
\newblock
\newblock
\urldef\tempurl%
\url{https://doi.org/10.48550/ARXIV.2207.00747}
\showDOI{\tempurl}


\bibitem[Wei et~al\mbox{.}(2022a)]%
        {wei2022emergent}
\bibfield{author}{\bibinfo{person}{Jason Wei}, \bibinfo{person}{Yi Tay},
  \bibinfo{person}{Rishi Bommasani}, \bibinfo{person}{Colin Raffel},
  \bibinfo{person}{Barret Zoph}, \bibinfo{person}{Sebastian Borgeaud},
  \bibinfo{person}{Dani Yogatama}, \bibinfo{person}{Maarten Bosma},
  \bibinfo{person}{Denny Zhou}, \bibinfo{person}{Donald Metzler},
  {et~al\mbox{.}}} \bibinfo{year}{2022}\natexlab{a}.
\newblock \showarticletitle{Emergent abilities of large language models}.
\newblock \bibinfo{journal}{\emph{arXiv preprint arXiv:2206.07682}}
  (\bibinfo{year}{2022}).
\newblock


\bibitem[Wei et~al\mbox{.}(2022b)]%
        {wei2022chainofthought}
\bibfield{author}{\bibinfo{person}{Jason Wei}, \bibinfo{person}{Xuezhi Wang},
  \bibinfo{person}{Dale Schuurmans}, \bibinfo{person}{Maarten Bosma},
  \bibinfo{person}{Brian Ichter}, \bibinfo{person}{Fei Xia},
  \bibinfo{person}{Ed Chi}, \bibinfo{person}{Quoc Le}, {and}
  \bibinfo{person}{Denny Zhou}.} \bibinfo{year}{2022}\natexlab{b}.
\newblock \bibinfo{title}{Chain of Thought Prompting Elicits Reasoning in Large
  Language Models}.
\newblock
\newblock
\urldef\tempurl%
\url{https://doi.org/10.48550/ARXIV.2201.11903}
\showDOI{\tempurl}


\bibitem[Wobbrock(2019)]%
        {wobbrock2019situationally}
\bibfield{author}{\bibinfo{person}{Jacob~O Wobbrock}.}
  \bibinfo{year}{2019}\natexlab{}.
\newblock \showarticletitle{Situationally aware mobile devices for overcoming
  situational impairments}. In \bibinfo{booktitle}{\emph{Proceedings of the ACM
  SIGCHI Symposium on Engineering Interactive Computing Systems}}.
  \bibinfo{pages}{1--18}.
\newblock


\bibitem[Wu et~al\mbox{.}(2021)]%
        {wu2021screen}
\bibfield{author}{\bibinfo{person}{Jason Wu}, \bibinfo{person}{Xiaoyi Zhang},
  \bibinfo{person}{Jeff Nichols}, {and} \bibinfo{person}{Jeffrey~P Bigham}.}
  \bibinfo{year}{2021}\natexlab{}.
\newblock \showarticletitle{Screen Parsing: Towards Reverse Engineering of UI
  Models from Screenshots}. In \bibinfo{booktitle}{\emph{The 34th Annual ACM
  Symposium on User Interface Software and Technology}} (Virtual Event, USA)
  \emph{(\bibinfo{series}{UIST '21})}. \bibinfo{publisher}{Association for
  Computing Machinery}, \bibinfo{address}{New York, NY, USA},
  \bibinfo{pages}{470–483}.
\newblock
\showISBNx{9781450386357}
\urldef\tempurl%
\url{https://doi.org/10.1145/3472749.3474763}
\showDOI{\tempurl}


\bibitem[Wu et~al\mbox{.}(2022)]%
        {wu2022aichain}
\bibfield{author}{\bibinfo{person}{Tongshuang Wu}, \bibinfo{person}{Michael
  Terry}, {and} \bibinfo{person}{Carrie~Jun Cai}.}
  \bibinfo{year}{2022}\natexlab{}.
\newblock \showarticletitle{AI Chains: Transparent and Controllable Human-AI
  Interaction by Chaining Large Language Model Prompts}. In
  \bibinfo{booktitle}{\emph{Proceedings of the 2022 CHI Conference on Human
  Factors in Computing Systems}} (New Orleans, LA, USA)
  \emph{(\bibinfo{series}{CHI '22})}. \bibinfo{publisher}{Association for
  Computing Machinery}, \bibinfo{address}{New York, NY, USA}, Article
  \bibinfo{articleno}{385}, \bibinfo{numpages}{22}~pages.
\newblock
\showISBNx{9781450391573}
\urldef\tempurl%
\url{https://doi.org/10.1145/3491102.3517582}
\showDOI{\tempurl}


\bibitem[Zaheer et~al\mbox{.}(2020)]%
        {lengthtransformer}
\bibfield{author}{\bibinfo{person}{Manzil Zaheer}, \bibinfo{person}{Guru
  Guruganesh}, \bibinfo{person}{Avinava Dubey}, \bibinfo{person}{Joshua
  Ainslie}, \bibinfo{person}{Chris Alberti}, \bibinfo{person}{Santiago
  Ontanon}, \bibinfo{person}{Philip Pham}, \bibinfo{person}{Anirudh Ravula},
  \bibinfo{person}{Qifan Wang}, \bibinfo{person}{Li Yang}, {and}
  \bibinfo{person}{Amr Ahmed}.} \bibinfo{year}{2020}\natexlab{}.
\newblock \showarticletitle{Big Bird: Transformers for Longer Sequences}.
\newblock  (\bibinfo{year}{2020}).
\newblock
\urldef\tempurl%
\url{https://doi.org/10.48550/ARXIV.2007.14062}
\showDOI{\tempurl}


\bibitem[Zhang et~al\mbox{.}(2021)]%
        {zhang2021screen}
\bibfield{author}{\bibinfo{person}{Xiaoyi Zhang}, \bibinfo{person}{Lilian de
  Greef}, \bibinfo{person}{Amanda Swearngin}, \bibinfo{person}{Samuel White},
  \bibinfo{person}{Kyle Murray}, \bibinfo{person}{Lisa Yu}, \bibinfo{person}{Qi
  Shan}, \bibinfo{person}{Jeffrey Nichols}, \bibinfo{person}{Jason Wu},
  \bibinfo{person}{Chris Fleizach}, \bibinfo{person}{Aaron Everitt}, {and}
  \bibinfo{person}{Jeffrey~P Bigham}.} \bibinfo{year}{2021}\natexlab{}.
\newblock \showarticletitle{Screen Recognition: Creating Accessibility Metadata
  for Mobile Applications from Pixels}. In
  \bibinfo{booktitle}{\emph{Proceedings of the 2021 CHI Conference on Human
  Factors in Computing Systems}} (Yokohama, Japan) \emph{(\bibinfo{series}{CHI
  '21})}. \bibinfo{publisher}{Association for Computing Machinery},
  \bibinfo{address}{New York, NY, USA}, Article \bibinfo{articleno}{275},
  \bibinfo{numpages}{15}~pages.
\newblock
\showISBNx{9781450380966}
\urldef\tempurl%
\url{https://doi.org/10.1145/3411764.3445186}
\showDOI{\tempurl}


\bibitem[Zhou et~al\mbox{.}(2022)]%
        {zhou2022leasttomost}
\bibfield{author}{\bibinfo{person}{Denny Zhou}, \bibinfo{person}{Nathanael
  Schärli}, \bibinfo{person}{Le Hou}, \bibinfo{person}{Jason Wei},
  \bibinfo{person}{Nathan Scales}, \bibinfo{person}{Xuezhi Wang},
  \bibinfo{person}{Dale Schuurmans}, \bibinfo{person}{Olivier Bousquet},
  \bibinfo{person}{Quoc Le}, {and} \bibinfo{person}{Ed Chi}.}
  \bibinfo{year}{2022}\natexlab{}.
\newblock \bibinfo{title}{Least-to-Most Prompting Enables Complex Reasoning in
  Large Language Models}.
\newblock
\newblock
\urldef\tempurl%
\url{https://doi.org/10.48550/ARXIV.2205.10625}
\showDOI{\tempurl}


\end{thebibliography}

\appendix

\section{Example Prompts for Feasibility Experiments Tasks}
We show the prompt examples used in each task from the feasibility experiments. For screen summarization, screen QA, and mapping instruction to UI action, we include 1-shot prompts for simplicity. For the screen question-generation study, we include the 2-shot prompt--one exemplar uses question combination (the second screen), and the other does not. Both exemplars contain a chain of thoughts consisting of intermediate results.

\subsection{Screen Question-Generation}
\label{prompt:1}
\begin{lstlisting}[language=html, caption=2-shot example prompt for screen question-generation.]
Given a screen, the agent needs to identify the elements requiring user input and generates corresponding questions.

Screen:
<p id=0 class="alertTitle">  Create password  </p>
<div id=1 class="titleDivider">   </div>
<input id=2 class="password">  Crowd3116  </input>
<input id=3 class="confirm password">  Crowd3116  </input>
<input id=4 class="hint">  c3  </input>
<input id=5 class="edEmailAddress">  appcrawler4@gmail.com  </input>
<p id=6 class="tvEmailAddressInfo">  This email address will be used to reset your password.  </p>
<button id=7 class="button2">  Cancel  </button>
<button id=8 class="button1">  OK  </button>

Now reasoning starts:
Q: How many input tags are there on the screen?
A: 4
Q: What is the purpose of the screen?
A: Create password.

It's a create password page and there are 4 input tags, including: 
1. id=2 asks for password.
2. id=3 asks to confirm password.
3. id=4 asks for hint.
4. id=5 asks for email address.

To help the user proceed with the screen, an agent will ask:
<SOQ>What password do you want to create? (id=2)<EOQ> 
<SOQ>Could you enter the password again to confirm? (id=3)<EOQ> 
<SOQ>What hint do you want to set? (id=4)<EOQ> 
<SOQ>What is your email in case you need to reset the password? (id=5)<EOQ>

Screen: 
<p id=0> IRS2Go,  </p>
<button id=1 alt="Open navigation drawer">  </button>
<p id=2 class="titleRefund"> Refund Status </p>
<p id=3 class="refundHeaderText"> Check your refund status by entering your information as shown on your 2015 tax return. This tool is updated no more than once every 24 hours, usually overnight.  </p>
<input id=4 class="taxId3Edit" alt="First 3 Digits of Social Security Number">  </input>
<p id=5 class="dash1"> - </p>
<input id=6 class="taxId2Edit" alt="Middle 2 Digits of Social Security Number">  </input>
<p id=7 class="dash2"> - </p>
<input id=8 class="taxId4Edit" alt="Last 4 Digits of Social Security Number">  </input>
<p id=9> Filing Status </p>
<input id=10 class="refundAmountEdit">  </input>
<button id=11 class="privacyNoticeButton" alt="Privacy Notice"> Privacy Notice,  </button>
<button id=12 class="getStatusButton" alt="Get Status"> GET STATUS,  </button>
<div id=13 class="navigationBarBackground">  </div>
<div id=14 class="statusBarBackground">  </div>

Now reasoning starts:
Q: How many input tags are there on the screen?
A: 4
Q: What is the purpose of the screen?
A: Check your refund status. 

It's a check refund status page and there are 4 input tags, including: 
1. id=4 asks for first 3 digits of SSN
2. id=6 asks for middle 2 digits of SSN
3. id=8 asks for last 4 digits of SSN
4. id=10 asks for the amount of refund.

To help the user proceed with the screen, an agent will ask:
<SOQ>What is your SSN? (id=4, id=6, id=8)<EOQ> 
<SOQ>What is the refund amount? (id=10)<EOQ> 
"""

\end{lstlisting}

\subsection{Screen Summarization}
\label{prompt:2}
\begin{lstlisting}[language=html, caption=1-shot example prompt for screen summarization.]
Given a screen, summarize its purpose.

Screen:
<img id=0>  </img>
<p id=1 class="cliv name textview"> Create new contact  </p>
<img id=2>  </img>
<p id=3 class="cliv name textview"> Add to a contact  </p>
<img id=4>  </img>
<p id=5 class="cliv name textview"> Send SMS  </p>
<button id=6 class="floating action button" alt="dial pad">  </button>
<button id=7 class="search back button" alt="stop searching">  </button>
<input id=8 class="search view"> 18773312998  </input>
<img id=9 class="search close button" alt="Clear search">  </img>
<div id=10 class="navigationBarBackground">  </div>
<div id=11 class="statusBarBackground">  </div>

Summary: <SOS>Screen of contact settings options<EOS>
\end{lstlisting}

\subsection{Screen Question-Answering}
\label{prompt:3}
\begin{lstlisting}[language=html, caption=1-shot example prompt for screen question-answering]
Given a mobile screen and a question, provide the answer based on the screen information.

Screen:
<p id=0> Invite for T20 Fans Live Chat  </p>
<button id=1 alt="Choose account">  </button>
<p id=2 class="menu send" alt="Send">  </p>
<p id=3 class="message header"> Message  </p>
<input id=4 class="message"> Join me on T20 Fans Live chat.  </input>
<div id=5 class="message separator">  </div>
<p id=6 class="message limit">  </p>
<div id=7 class="separator">  </div>
<p id=8 class="selection"> Add recipients  </p>
<div id=9 class="separator">  </div>
<p id=10 class="text"> Suggestions from Google  </p>
<p id=11> A,  </p>
<p id=12 class="name"> appcrawler5@gmail.com  </p>
<p id=13 class="contact detail"> appcrawler5@gmail.com  </p>
<img id=14 class="contact method">  </img>
<div id=15 class="divider">  </div>
<p id=16> A,  </p>
<p id=17 class="name"> appcrawler4@gmail.com  </p>
<p id=18 class="contact detail"> appcrawler4@gmail.com  </p>
<img id=19 class="contact method">  </img>
<div id=20 class="divider">  </div>
<p id=21 class="text"> Everyone  </p>
<img id=22>  </img>
<p id=23 class="name"> App Crawler  </p>
<p id=24 class="contact detail"> (415) 336-5454  </p>
<img id=25 class="contact method">  </img>
<img id=26 class="channel switcher icon">  </img>
<div id=27 class="divider">  </div>
<p id=28> T,  </p>
<p id=29 class="name"> test,  </p>
<p id=30 class="contact detail"> (415) 336-5454  </p>
<img id=31 class="contact method">  </img>
<img id=32 class="channel switcher icon">  </img>
<div id=33 class="divider">  </div>
<div id=34 class="navigationBarBackground">  </div>
<div id=35 class="statusBarBackground">  </div>

Q: What email addresses are there?
A: <SOA>appcrawler5@gmail.com<EOA>
\end{lstlisting}

\subsection{Mapping Instruction to UI Action}
\label{prompt:4}
\begin{lstlisting}[language=html, caption=1-shot example prompt for mapping instruction to UI action.]
Given a screen, an instruction, predict the id of the UI element to perform the instruction.

Screen:
<div id=0 alt="Apps list">  </div>
<img id=1 class="g icon">  </img>
<img id=2 class="mic icon" alt="Voice search">  </img>
<p id=3 class="icon" alt="Calculator"> Calculator  </p>
<p id=4 class="icon" alt="Calendar"> Calendar  </p>
<p id=5 class="icon" alt="Camera"> Camera  </p>
<p id=6 class="icon" alt="Chrome"> Chrome  </p>
<p id=7 class="icon" alt="Clock"> Clock  </p>
<p id=8 class="icon" alt="Contacts"> Contacts  </p>
<p id=9 class="icon" alt="Custom Locale"> Custom Locale  </p>
<p id=10 class="icon" alt="Dev Tools"> Dev Tools  </p>
<p id=11 class="icon" alt="Drive"> Drive  </p>
<p id=12 class="icon" alt="Files"> Files  </p>
<p id=13 class="icon" alt="Gmail"> Gmail  </p>
<p id=14 class="icon" alt="Google"> Google  </p>
<p id=15 class="icon" alt="Hangouts"> Hangouts  </p>
<p id=16 class="icon" alt="Maps"> Maps  </p>
<p id=17 class="icon" alt="Messages"> Messages  </p>
<p id=18 class="icon" alt="Phone"> Phone  </p>
<p id=19 class="icon" alt="Photos"> Photos  </p>
<p id=20 class="icon" alt="Play Movies & TV"> Play Movies & TV  </p>
<p id=21 class="icon" alt="Play Music"> Play Music  </p>
<p id=22 class="icon" alt="Settings"> Settings  </p>
<p id=23 class="icon" alt="WebView Browser Tester"> WebView Browser Tester  </p>
<p id=24 class="icon" alt="YouTube"> YouTube  </p>
<p id=25 class="icon" alt="Photos"> Photos  </p>
<p id=26 class="icon" alt="Maps"> Maps  </p>
<p id=27 class="icon" alt="Contacts"> Contacts  </p>
<p id=28 class="icon" alt="Settings"> Settings  </p>
<p id=29 class="icon" alt="Clock"> Clock  </p>
<div id=30 class="fast scroller">  </div>
<div id=31>  </div>
<div id=32 class="hotseat">  </div>

Instruction: Open your device's Clock app.
Prediction: id=<SOI>29<EOI>

\end{lstlisting}






\end{document}